\documentclass[fleqn,useAMS,usenatbib]{mnras}

\usepackage{graphicx, hyperref, aas}
\usepackage{float}
\usepackage[T1]{fontenc}
\usepackage{ae,aecompl}
\usepackage{amsmath}	
\usepackage{amssymb}	

\def\aap{Astronomy \& Astrophysics}
\def\apj{The Astrophysical Journal}
\def\mnras{Monthly Notices of the Royal Astronomical Society}

\title[Polarization leakage in wide fields]{Polarization leakage in epoch of reionization windows --\\ III. Wide-field effects of narrow-field arrays}
\author[K. M. B. Asad et al.]
{K. M. B. Asad$^{1,2,3,4}$\thanks{E-mail: khmbasad@gmail.com},
L. V. E. Koopmans$^{4}$,
V. Jeli\'{c}$^{5,4,6}$,
A. G. de Bruyn$^{4,6}$,
\newauthor
V. N. Pandey$^6$ and
B. K. Gehlot$^4$ \\
$^1$Department of Physics, University of the Western Cape, Cape Town 7535, South Africa\\
$^2$Department of Physics and Electronics, Rhodes University, PO
Box 94, Grahamstown, 6140, South Africa\\
$^3$SKA South Africa, 3rd Floor, The Park, Park Road, Pinelands, 7405 South Africa\\
$^4$Kapteyn Astronomical Institute, University of Groningen, PO Box 800, NL-9700 AV Groningen, the Netherlands\\
$^5$Ru{\dj}er Bo\v{s}kovi\'{c} Institute, Bijeni\v{c}ka cesta 54, 10000 Zagreb, Croatia\\
$^6$ASTRON, PO Box 2, NL-7990 AA Dwingeloo, the Netherlands\\
}

\begin{document}

\date{Accepted . Received ; in original form}

\pagerange{\pageref{firstpage}--\pageref{lastpage}} \pubyear{2017}

\maketitle

\label{firstpage}

\begin{abstract}
Leakage of polarized Galactic diffuse emission into total intensity can potentially mimic the 21-cm signal coming from the epoch of reionization (EoR), as both of them might have fluctuating spectral structure.
Although we are sensitive to the EoR signal only in small fields of view, chromatic sidelobes from further away can contaminate the inner region.
Here, we explore the effects of leakage into the `EoR window' of the cylindrically averaged power spectra (PS) within wide fields of view using both observation and simulation of the 3C196 and NCP fields, two observing fields of the LOFAR-EoR project.
We present the polarization PS of two one-night observations of the two fields and find that the NCP field has higher fluctuations along frequency, and consequently exhibits more power at high-$k_\parallel$ that could potentially leak to Stokes $I$.
Subsequently, we simulate LOFAR observations of Galactic diffuse polarized emission based on a model to assess what fraction of polarized power leaks into Stokes $I$ because of the primary beam.
We find that the rms fractional leakage over the instrumental $k$-space is $0.35\%$ in the 3C196 field and $0.27\%$ in the NCP field, and it does not change significantly within the diameters of $15^\circ$, $9^\circ$ and $4^\circ$.
Based on the observed PS and simulated fractional leakage, we show that a similar level of leakage into Stokes $I$ is expected in the 3C196 and NCP fields, and the leakage can be considered to be a bias in the PS.
\end{abstract}


\section{Introduction}
Polarization leakage is one of the least explored effects that can potentially contaminate the 21-cm signal coming from the epoch of reionization (EoR).
A fraction of the polarized emission (Stokes $Q$, $U$) always leaks into total intensity (Stokes $I$) due to instrumental effects, specifically a mismatch of the  primary beams (PB) of the two feeds of an antenna.
The EoR signal is expected to be detected statistically by current telescopes such as GMRT \citep{Paciga2011}, LOFAR \citep{vh13}, MWA \citep{Tingay2013} and PAPER \citep{Parsons2010}, and future telescopes such as HERA \citep{Deboer2017} and SKA \citep{Koopmans2015}.
For a successful detection, the foregrounds contaminating the signal need to be removed one by one.
First, bright point sources are removed.
Then, the total intensity of the Galactic diffuse synchrotron emission is removed by utilizing the fact that this emission is spectrally smooth, whereas the EoR signal is not \citep{je08,da10,ha10,tr12,mo12,be13,po13,ch13,di15,th15}.
Removing polarization leakage comes at the very end, if at all necessary, because the leakage level is lower than noise in the current observations \citep{as15,ko16}, although potentially still above the EoR signal.

\citet{je10} showed that if the polarization angle of the Galactic diffuse polarized emission is differentially Faraday-rotated by the magnetized plasma in the interstellar medium, the emission that reaches us can have significant spectral fluctuations.
If this is indeed the case, and if this high-rotation measure polarized emission is leaked into Stokes $I$, the leakage might mimic the EoR signal which is expected to have similar fluctuations as a function of frequency.

There are two main approaches toward detecting the EoR signal---foreground `avoidance' and `removal' \citep{ch14}.
In the former approach, the region of the cylindrically averaged power spectra (PS) most contaminated by foregrounds and noise is avoided.
The region least contaminated by the foregrounds and systematics is called the `EoR window' and this is the only region where the EoR signal is looked for.
In the latter approach, foregrounds are supposed to be subtracted from the data employing various strategies.

The levels of foreground and system noise are expected to be much higher than the polarization leakage in Stokes $I$.
For example, in the LOFAR-EoR observations, an excess noise is detected, which is higher than the expected level of leakage, and this noise is not contributed by leakage \citep{Patil2017}.
In case of PAPER, \citet{ko16} found no evidence of polarization leakage in the EoR window with their current sensitivity.
Although polarization leakage is not one of the main concerns of the current EoR experiments, previously it was thought to pose a greater problem.
For example, based on the experience of WSRT\footnote{Westerbork Synthesis Radio Telescope, \url{http://www.astron.nl/radio-observatory/astronomers/wsrt-astronomers}.}, a higher level of polarization leakage was expected than what was found by \citet{as15} in case of LOFAR.
The diffuse polarized emission has been found to be rich in Faraday structures in different fields \citep{Iacobelli13, je14, je15, VanEck17}, but the instrumental polarization of LOFAR is much lower than WSRT resulting in a lower leakage in the LOFAR observations.
Although the level of leakage is low, it is worth exploring because it will be relevant for the more sensitive experiments in the future.
Once we reach the sensitivity limit where leakage becomes relevant, we have to decide whether to avoid or remove leakage.
If the leakage toward certain directions is found to be spectrally structured avoidance would not be a proper strategy.

\citet{as15} showed that in the 3C196 field, one of the observing fields of the LOFAR-EoR project, leakage from observed polarized emission into Stokes $I$ is contained within a wedge-shaped region at the high-$k_\perp$ (transverse wavenumber), low-$k_\parallel$ (line of sight wavenumber) corner of the cylindrical PS, and there is a wide region at the opposite corner of the PS relatively free from leakage-contamination.
This prediction of leakage was performed using the model PB of LOFAR.
\citet{as16} showed that this PB model has an accuracy of $\sim 10\%$ within the first null of the PB.
These two papers enabled us to properly assess the level of leakage that could potentially make EoR detection harder, if not mitigated properly.

In this paper, we take our previous analyses one step further.
Here, we present cylindrical PS of both the 3C196 and NCP fields and compare them to each other.
Another worry about polarization leakage is that the leakage is expected to increase with distance from the phase center---wider fields are expected to suffer from more leakage.
Although in case of LOFAR, we are sensitive to the EoR signal only within small fields of view (FoV), chromatic sidelobes of the diffuse and compact emission and the polarization leakage from further away might corrupt the inner regions to some extent.
We therefore present simulations of leakage for different FoV ranging from $4^\circ\times 4^\circ$ to $15^\circ\times 15^\circ$, and compare their PS.
Once we know the fractional leakage, we can predict the level of leakage based on the observed diffuse polarized emission in different fields.

This paper is organized as follows: Section \ref{s:pspe} presents the PS of the observed diffuse polarized emission in the 3C196 and NCP fields within \textit{two} different FoV.
Section \ref{s:flwf} presents the calculation of the fractional leakage, by predicting LOFAR observations of a simulated Galactic diffuse polarized emission, for the two fields and for three different FoV.
In both sections, the methods and results are presented in separate subsections.
The implications of the calculated fractional leakage in Section \ref{s:flwf} on the PS, presented in Section \ref{s:pspe}, are explored in the discussion section.
The paper ends with the conclusion and some remarks about our ongoing and future works.

\section{Power spectra of polarized emission} \label{s:pspe}
This section presents cylindrically averaged two-dimensional (2D) PS of the diffuse polarized emission in the 3C196 \citep{je15} and NCP observing fields.
A PS of the polarized emission $P=Q+iU$ in the 3C196 field was shown in \citet[hereafter A15]{as15}, but here we present the original Stokes $Q$ and $U$ PS.
Moreover, the PS of A15 were created from the observed emission convolved with a PB model, whereas here we create PS from the observed emission itself.
In the previous case, the PS was, in effect, convolved with the PB twice, as the observed data was not deconvovled to correct for the PB before convolving it again with the PB.
This double-convolution was not that relevant because in that test, our aim was to calculate the \textit{fractional} polarization leakage.
We found that the ratio of the polarized emission and leakage is almost constant over the instrumental $k$-space.
It is indeed expected to be constant, because leakage is caused by a convolution of the visibilities with a polarized PB that remains \textit{almost} the same for all baselines, and thus at all $k$-scales.
Therefore, if we know the PS of the polarized emission and the fractional leakage for a certain field, we can predict the level of leakage in $k$-space and potentially correct for this bias in the EoR PS, in a similar fashion as the correction for the noise bias.
The PS of polarization in the 3C196 and NCP fields presented in this paper provide an idea of how much the `EoR window' can potentially be corrupted toward different pointing directions.

\begin{table}
\centering
\begin{minipage}{\linewidth}
\centering
\caption{Parameters used for power spectrum estimation:}
\label{t:3setup}
\begin{tabular}{@{}lllr@{}}
\hline
\hline
\\
& NCP & 3C196 \\
\cline{2-3}\\
Observation ID & L86762 & L80508 \\
Phase center ($\alpha,\delta$) & 0$^\circ$, 90$^\circ$ & 123.4$^\circ$, 48.2$^\circ$ \\
Observation date & 6 February 2013 & 16 Dec 2012 \\
Observing time & 13 hours & 8 hours \\
Baseline range & \multicolumn{2}{c}{30--800 $\lambda$} \\
Frequency range & \multicolumn{2}{c}{150--160 MHz} \\
Spectral subbands & \multicolumn{2}{c}{50} \\
Bandwidth & \multicolumn{2}{c}{10 MHz} \\
Spectral resolution & \multicolumn{2}{c}{0.195 MHz} \\
Integration time & \multicolumn{2}{c}{10 s} \\
Pixel scale & \multicolumn{2}{c}{0.5 arcmin} \\
\\
\hline
\hline
\end{tabular}
\end{minipage}
\end{table}

\begin{figure*}
\centering
\includegraphics[width=0.9\linewidth]{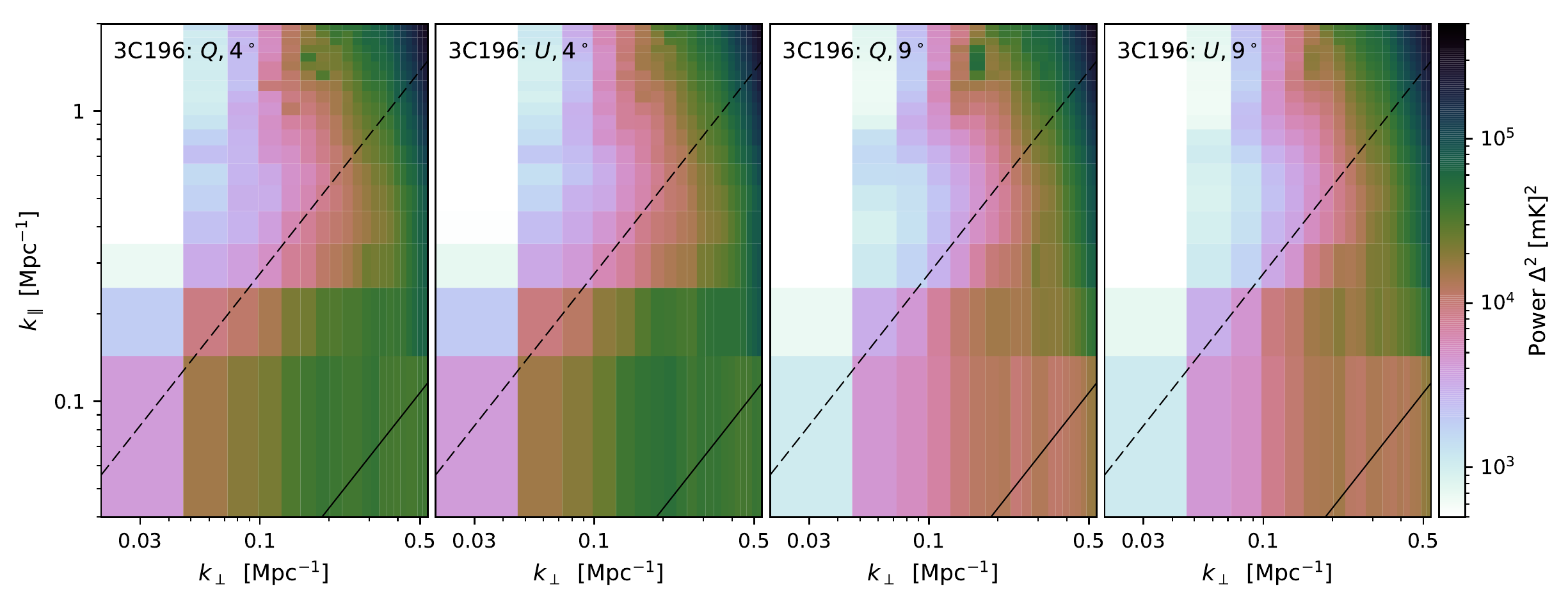}

\includegraphics[width=0.9\linewidth]{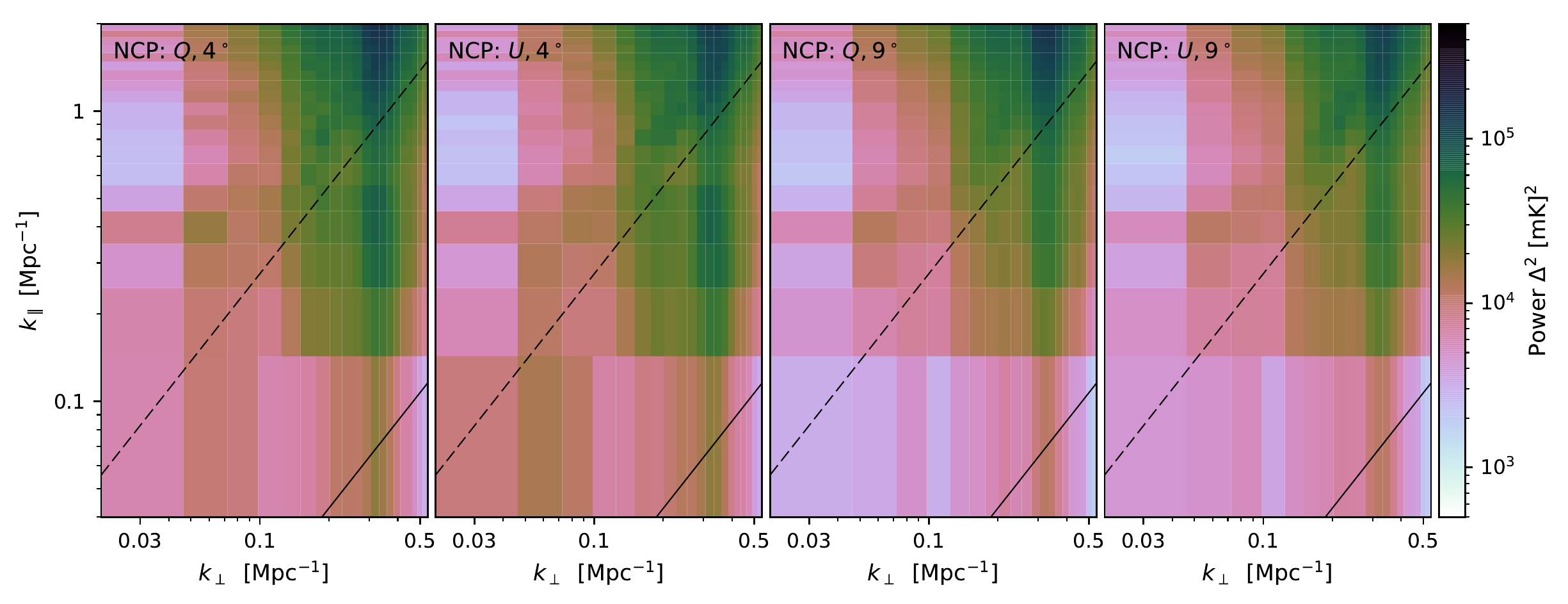}
\caption{Cylindrically averaged 2D power spectra of the polarized emission observed by LOFAR in the inner $4^\circ\times 4^\circ$ (first two columns from the left) and $9^\circ\times 9^\circ$ (last two columns) of the 3C196 (\textit{top row}) and NCP (\textit{bottom row}) fields within the frequency range of 150--160 MHz. Both Stokes $Q$ (first and third columns) and $U$ (second and fourth columns) spectra are shown in units of [mK]$^2$. The lower solid and upper dashed lines correspond to the boundaries of the primary and horizon wedges respectively.}
\label{f:ps2d}
\end{figure*}

\begin{figure*}
\begin{minipage}[b]{0.45\linewidth}
\includegraphics[width=\linewidth]{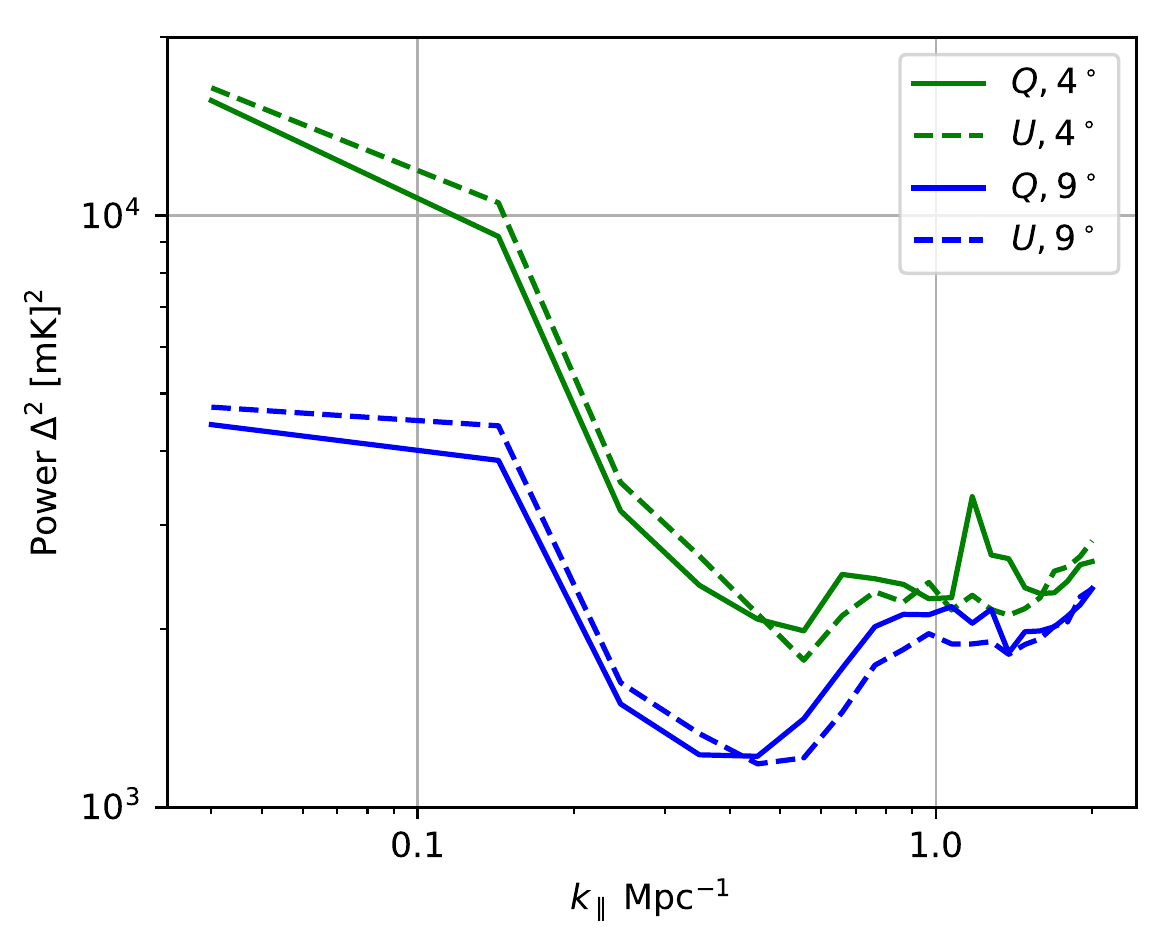}
\end{minipage}
\begin{minipage}[b]{0.45\linewidth}
\includegraphics[width=\linewidth]{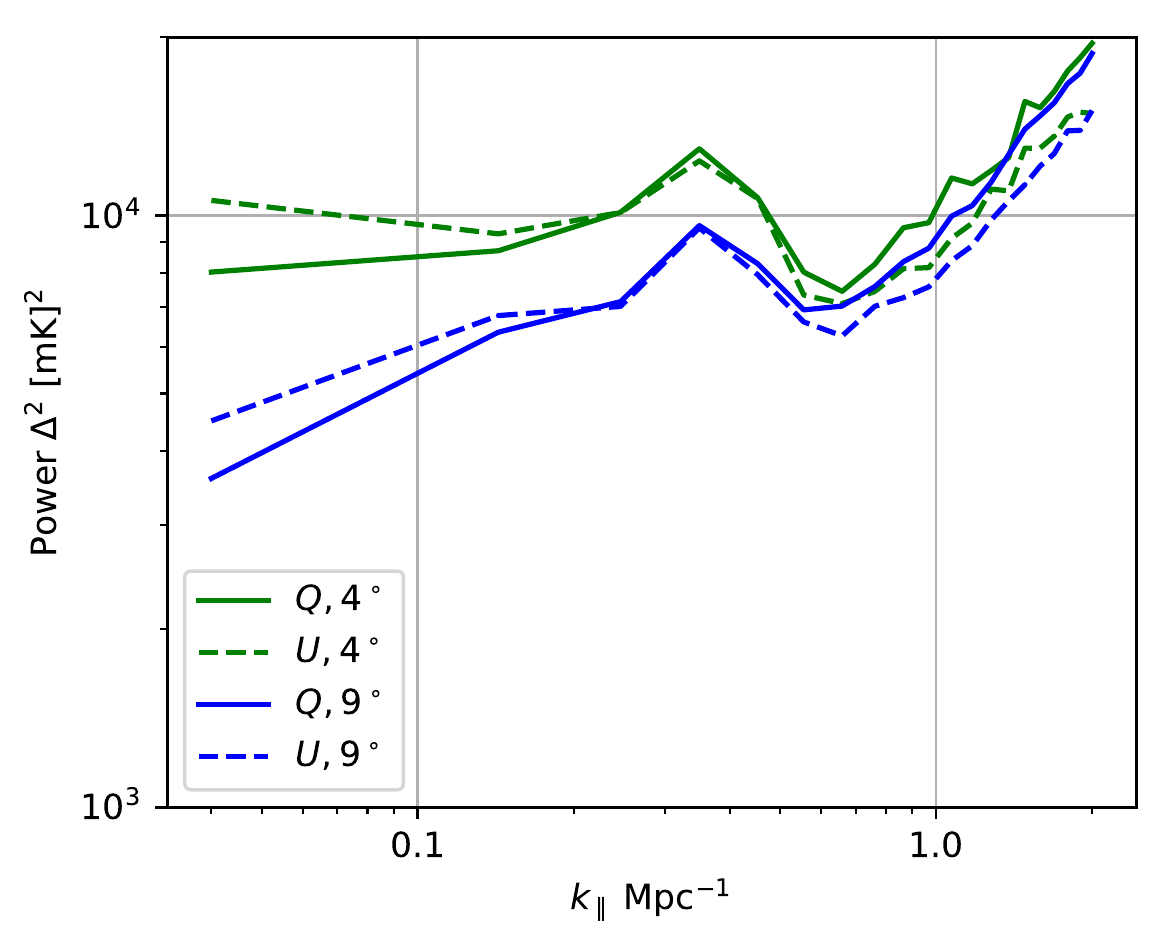}
\end{minipage}
\caption{Power spectra of polarized emission observed by LOFAR as a function of $k_\parallel$ averaged over $k_\perp\sim 0.02 - 0.1$ Mpc$^{-1}$ in the 3C196 (\textit{left}) and NCP (\textit{right}) fields within $4^\circ\times 4^\circ$ (green) and $9^\circ\times 9^\circ$ (blue).
The solid and dashed lines correspond to the Stokes $Q$ and $U$ spectra, respectively.
}
\label{f:ps1d}
\end{figure*}

\subsection{Method}
The PS have been created using the same pipeline as used in A15.
Here, we briefly outline the main steps relevant for this section.
We took polarization images (Stokes $Q$ and $U$) of 50 spectral subbands corresponding to about 10 MHz of bandwidth from 150 to 160 MHz for both the NCP and 3C196 fields.
The images are created from the observed visibilities, calibrated using the \textit{full-polarization direction-independent self-calibration} package {\tt BBS}, using `natural' weighting and an angular resolution of 0.5 arcmin within a diameter of $10^\circ$.
Two sets of image cubes are produced from these images, one containing the inner $4^\circ$, and the other the inner $9^\circ$ in diameter.
These two diameters were chosen because it will make the comparison of these PS with the PS from the exercise described in the next section more convenient.
The image cubes are then Fourier transformed, and the squared absolute values of the transformed cubes are cylindrically averaged resulting in a 2D PS as a function of comoving transverse ($k_\perp$) and line-of-sight ($k_\parallel$) wavenumbers.
Finally, the power is averaged over some $k_\perp$ values to create an one-dimensional PS as a function of $k_\parallel$.

In this papaer, an additional feature, the \textit{wedge}, has been indicated in the 2D PS.
The total intensity of diffuse foregrounds is expected to be smooth along frequency, and hence should appear only at low $k_\parallel$.
However, the frequency-dependent synthesized beam (PSF) leaks power from low $k_\parallel$ to high $k_\parallel$ as one goes to higher $k_\perp$ wavenumbers, spreading the foreground in a wedge-shaped region on the bottom-right corner of a PS.
The spread increases at higher $k_\perp$ because longer baselines have higher fringe rates making them more susceptible to spectral distortions.
In an optimistic scenario, the wedge should spread no further than the region delimited by the FWHM of the PB (beamwidth) of an instrument.
In fact, the line delimiting the foreground wedge is calculated from the beamwidth ($\theta_{\tt FoV}$) as \citep{mo12,li14a,di14}
\begin{equation}
k_\parallel = \left[\sin\theta_{\tt FoV} \frac{H_0D_c(z)E(z)}{c(1+z)}\right] k_\perp
\end{equation}
where $H_0$ is the Hubble parameter at redshift $z=0$, $c$ is the speed of light, $E(z)$ is the dimensionless Hubble parameter defined in A15, and $D_c(z)\equiv\int_0^z dz'/E(z')$.
This wedge can be called the `primary wedge' as its boundary is fixed by the width of the PB.
In a pessimistic scenario, foreground might leak beyond the primary wedge, but even then it should not leak beyond the `horizon wedge', the wedge delimited by the angular distance to the horizon.
The region above the wedge is considered to be the `EoR window' as it is expected to be least contaminated by foreground and systematics.
Although the concept of wedge was originally developed for analyzing the Stokes $I$ PS \citep{Datta2010,ve12,pa12}, it can also be used for polarization PS, which we intend to do here.

\subsection{Results}\label{s:obs}

Fig. \ref{f:ps2d} shows 2D PS of the observed polarized diffuse emission (Stokes $Q$ and $U$) in the 3C196 (top row) and NCP (bottom row) fields within a diameter of $4^\circ$ (first two columns) and $9^\circ$ (last two columns).
The corresponding 1D PS, as a function of $k_\parallel$ averaged over $k_\perp\sim 0.02 - 0.1$ Mpc$^{-1}$, are shown in Fig. \ref{f:ps1d}.
These particular $k_\perp$ bins were used because the diffuse emission is more dominant at these scales.
In the cylindrical PS, the lower solid and upper dashed lines correspond to the boundaries of the `primary' and `horizon' wedges respectively.

The diffuse polarized emission in the 3C196 field is clearly contained within a wedge-shaped region, but the wedge extends beyond the limit set by the PB.
Significant power is seen between the primary and horizon wedges.
The region at $k_\parallel\ge 0.2$ and $k_\perp\le 0.15$ is relatively free of polarized emission and noise.
The power of the polarized emission can be seen more clearly from the corresponding 1D PS shown on the left panel of Fig. \ref{f:ps1d}.
The power of the diffuse emission is $\sim 10^4$ [mK]$^2$, and there is a drop of almost one order of magnitude at higher $k_\parallel$ scales.

The bottom panels of Fig. \ref{f:ps2d} show the PS of the observed polarized emission in the NCP field.
The diffuse emission is lower here compared to the 3C196 field, which is seen by comparing the low-$k_\parallel$ power in the top and bottom panels.
Another difference is that there is significant power at high $k_\parallel$ and low $k_\perp$ which will leak into the EoR window of the Stokes $I$ PS.
The difference is more obvious in the 1D PS---there is no drop of power at higher $k_\parallel$ scales in the right panel of Fig. \ref{f:ps1d}.
The power at high $k_\parallel$ is caused by differential Faraday rotation of the intrinsically polarized signal by the intervening magnetized plasma along the LOS.
The difference between the two fields, thus, shows that one can expect considerable difference in the level of Faraday rotation toward different directions in the sky, and hence considerable difference in the level of polarization leakage.

The power of polarized emission decreases in both the 3C196 and NCP fields, if we create PS from $9^\circ\times 9^\circ$, instead of $4^\circ\times 4^\circ$, which can be seen by comparing the corresponding plots in Fig. \ref{f:ps2d} and \ref{f:ps1d}.
This is primarily because of the fact that the region outside $4^\circ$ is dominated by noise, and averaging signals with noise within the larger area gives a lower power.
Averaging signals with anti-correlated position angles could be another potential reason, because this effect could be higher within the larger area.

These PS show that the `EoR window' could be more prone to leakage-contamination in the NCP field, if the level of fractional leakage in this field is comparable to the 3C196 field.
If, on the other hand, the fractional leakage is lower in the NCP field, a smaller fraction of the high level of power at high $k_\parallel$ would leak into Stokes $I$, making this field more or less similar to the 3C196 field.
We calculate the fractional leakage caused by the LOFAR PB model in the two fields in the next section.
The implications of the level of fractional leakage on the `EoR window' will be described in Section 4.

\section{Fractional leakage in wide fields} \label{s:flwf}
We previously found that the ratio between the power spectra of the polarized emission within $3\times 3$ degrees and its leakage into Stokes $I$ varies very little over the instrumental $k$-space (A15).
However, that ratio was calculated using a noisy polarization observation.
To understand this ratio better, and to see how it changes if the observing area is increased, we now perform a different experiment where a simulated model of polarized emission is used instead of an observation.
There are two interesting effects that one can show by simulating wider fields: the attenuation of polarized emission with distance from the phase center, and the chromatic side-lobes of leaked polarized emission near the nulls, where the PB is not very smooth in frequency.
For the purpose of this section, we are only interested in their overall impact on the rms fractional leakage.

\subsection{Method}
We have created a model of the diffuse Galactic polarized emission using the foreground simulations of \citet{je10}.
The model, described in Section \ref{s:model}, contains diffuse polarized foreground within $15^\circ\times 15^\circ$ and 150--160 MHz, and has a pixel scale of 18.75 arcmin, sufficient to analyze the short baselines we consider here.
A coarse grid of $50\times 50$ pixels is used because simulating visibilities for more pixels is computationally very expensive.
We note however that for the purpose of calculating polarization leakage, oversampling the PSF is not necessary.
The polarization model has both Stokes $Q$ and $U$ emission.
For each Stokes parameter three image cubes are created from the original model cube for three different observing areas: $15^\circ\times 15^\circ$, $9^\circ\times 9^\circ$, and $4^\circ\times 4^\circ$.
Hence, the three cubes contain $50^2$, $28^2$, and $16^2$ pixels, respectively, in the spatial domain.
The third (frequency) dimension is 50 in each case.

A point source sky model has to be created from this diffuse emission, because our simulation software can only handle point sources.
Therefore, each voxel in the Stokes $Q,U$ cubes is considered a point source with $I=V=0$ and $Q,U$ taken from the value of the voxel.
The three sky models for the three different diameters finally contained 2500, 784, and 256 polarized point sources.
Visibilities corresponding to these sky models are simulated toward two phase centers, one centered on the NCP, and the other on 3C 196.
All parameters for the six simulated observations, for the three diameters centered on two different directions, were kept same except for the total observing time.
Because NCP always remains above the horizon as seen from the LOFAR site, it can be observed for considerably more time than the 3C196 field, which in general can be oberved at high elevation only during the winter and spring seasons at fixed sidereal times, allowing typically 6h syntheses. To reconstruct the effects of the actual observations as precisely as possible, we have simulated the NCP observations for more time than 3C196.

LOFAR-EoR visibilities are usually integrated every 2 seconds, and the 2 s visibilities are again averaged down to every 10 s.
If, to remain as close to reality as possible, we wanted to simulate observations every 10 s for all LOFAR baselines, we would have to predict 442 million visibilities for 13 hours and for 50 spectral subbands, which is computationally very expensive.
Instead, we have predicted visibilities for every 120 s (as mentioned in Table \ref{t:3model}), and for only the core baselines of LOFAR, the longest baseline being only 3 km.
In the end, baselines longer than 180 $\lambda$ will not be needed, because the model has a resolution of only 18.75 arcmin, and any baseline longer than 180 $\lambda$ would resolve the diffuse emission into point sources.
Also note that time-smearing on these short baselines is negligible.

Full-Stokes visibilities are simulated in four main steps.
First, the sky model is Fourier transformed to produce four correlations of the visibilities for every baseline, frequency channel, and timeslot.
Then, the visibilities are convolved with the Fourier transform of the PB corresponding to the specified baseline, frequency, and timeslot, and toward the specified directions.
Third, the convolved visibilities are corrected for the polarized PB toward the phase center only via an inverse Mueller matrix multiplication.
After this correction, all effects of the PB toward the phase center are removed, and only the `differential' effects, i. e. those of the wide-field PB with respect to the phase center, remain.
Finally, Stokes visibilities are calculated from the four visibility correlations.
The first three steps are performed using the standard LOFAR calibration and simulation software {\tt BBS} \citep{pa09}.

The sources in the sky model does not have any Stokes $I$ flux, but Stokes $Q,U$ emission are leaked into Stokes $I,V$ resulting in non-zero values of the latters.
The aim of this exercise is to measure the fraction of power leaked from $Q$ and $U$ to $I$ because of the polarized PB, and in this respect it is similar to A15.
But there are two major differences.
First, in A15, fractional leakage was calculated from real observations, and here we calculate it from simulated observations to avoid the effect of noise and the leakage of Stokes $I$ into $Q,U$.
Second, the visibilities were predicted using {\tt AWIMAGER} in A15, whereas here {\tt BBS} is used.
In the previous case, a gridded sky map was directly Fourier transformed to produce visibilities and then convolved with the PB.
Here, direction-dependent Jones matrices corresponding to the PB model are applied to each source in the sky model.
Although {\tt AWIMAGER} is much faster, {\tt BBS} is used for the current exercise because currently {\tt AWIMAGER} cannot produce visibilities from maps wider than the main lobe of the PB.

\begin{figure*}
\centering
\begin{minipage}[b]{0.35\linewidth}
\includegraphics[width=\linewidth]{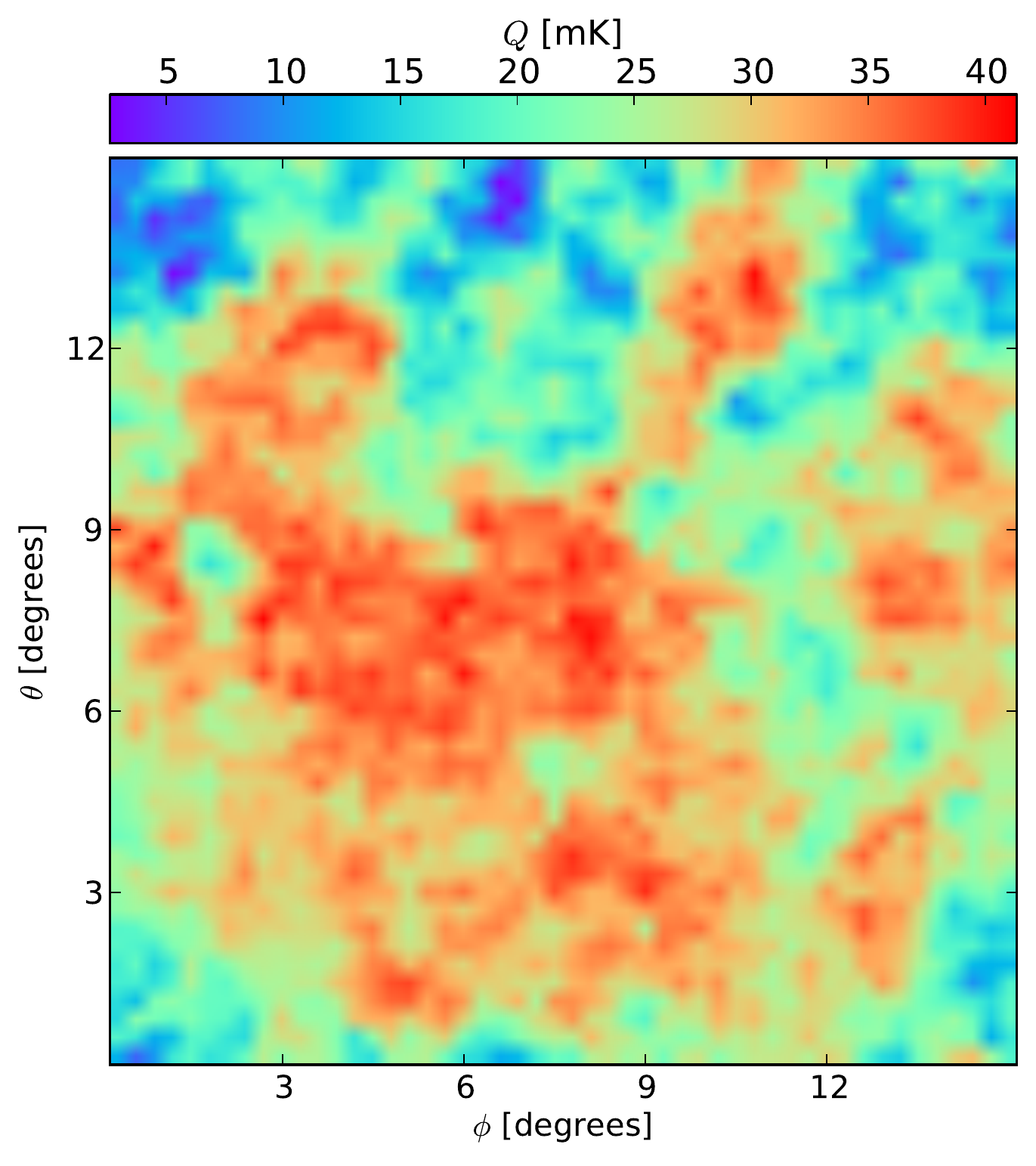}
\end{minipage}
\begin{minipage}[b]{0.35\linewidth}
\includegraphics[width=\linewidth]{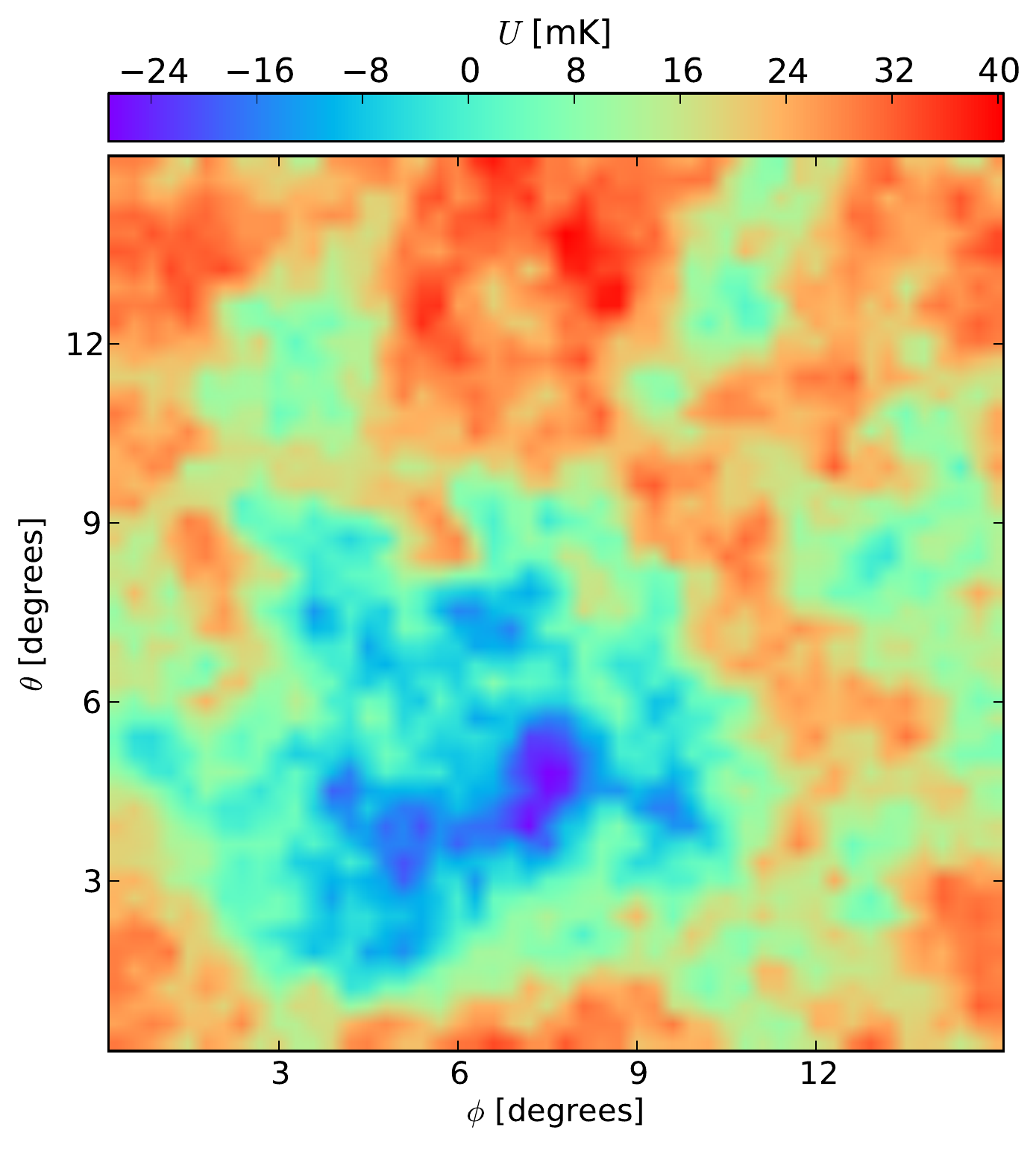}
\end{minipage}

\begin{minipage}[b]{0.36\linewidth}
\includegraphics[width=\linewidth]{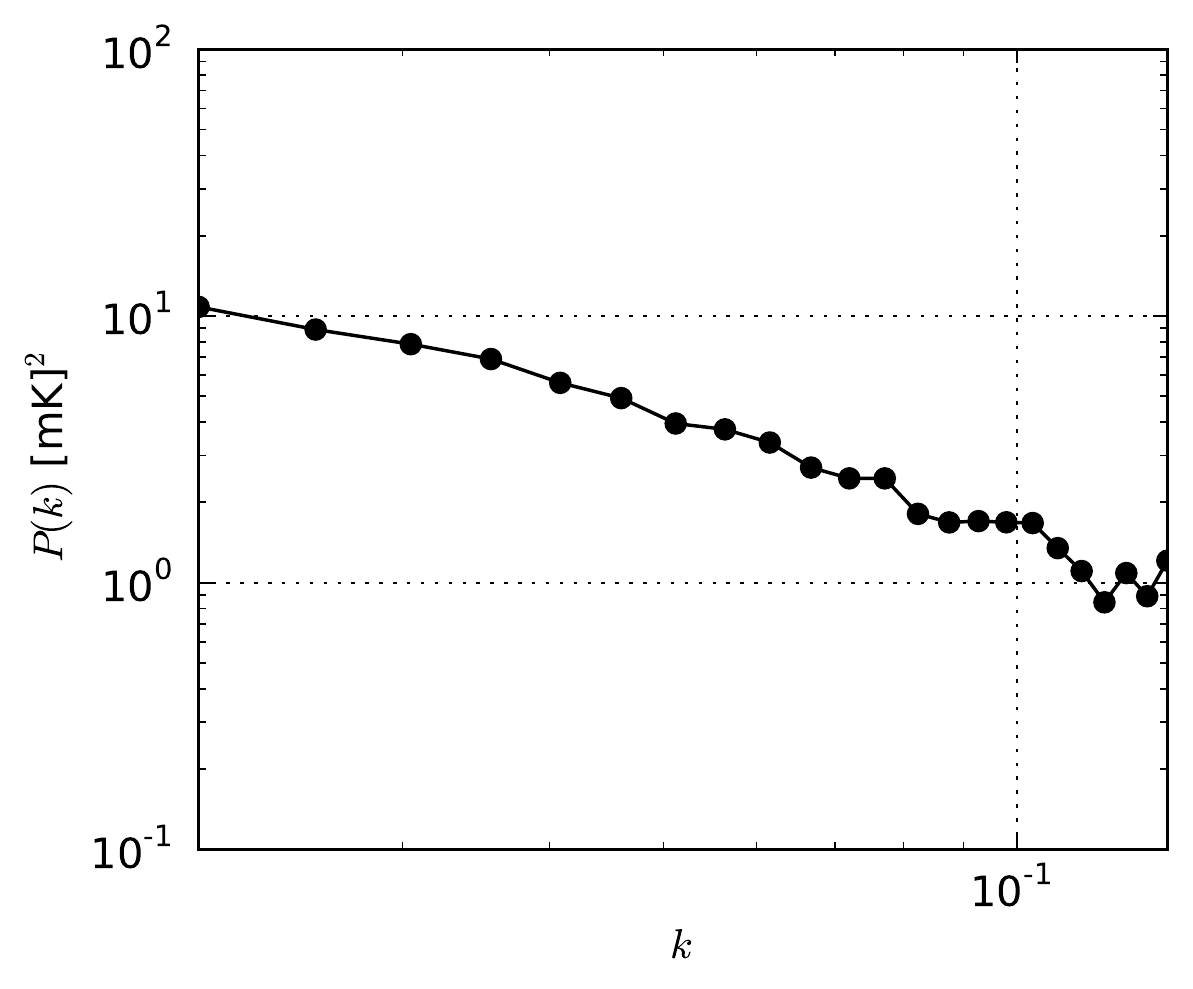}
\end{minipage}
\begin{minipage}[b]{0.38\linewidth}
\includegraphics[width=\linewidth]{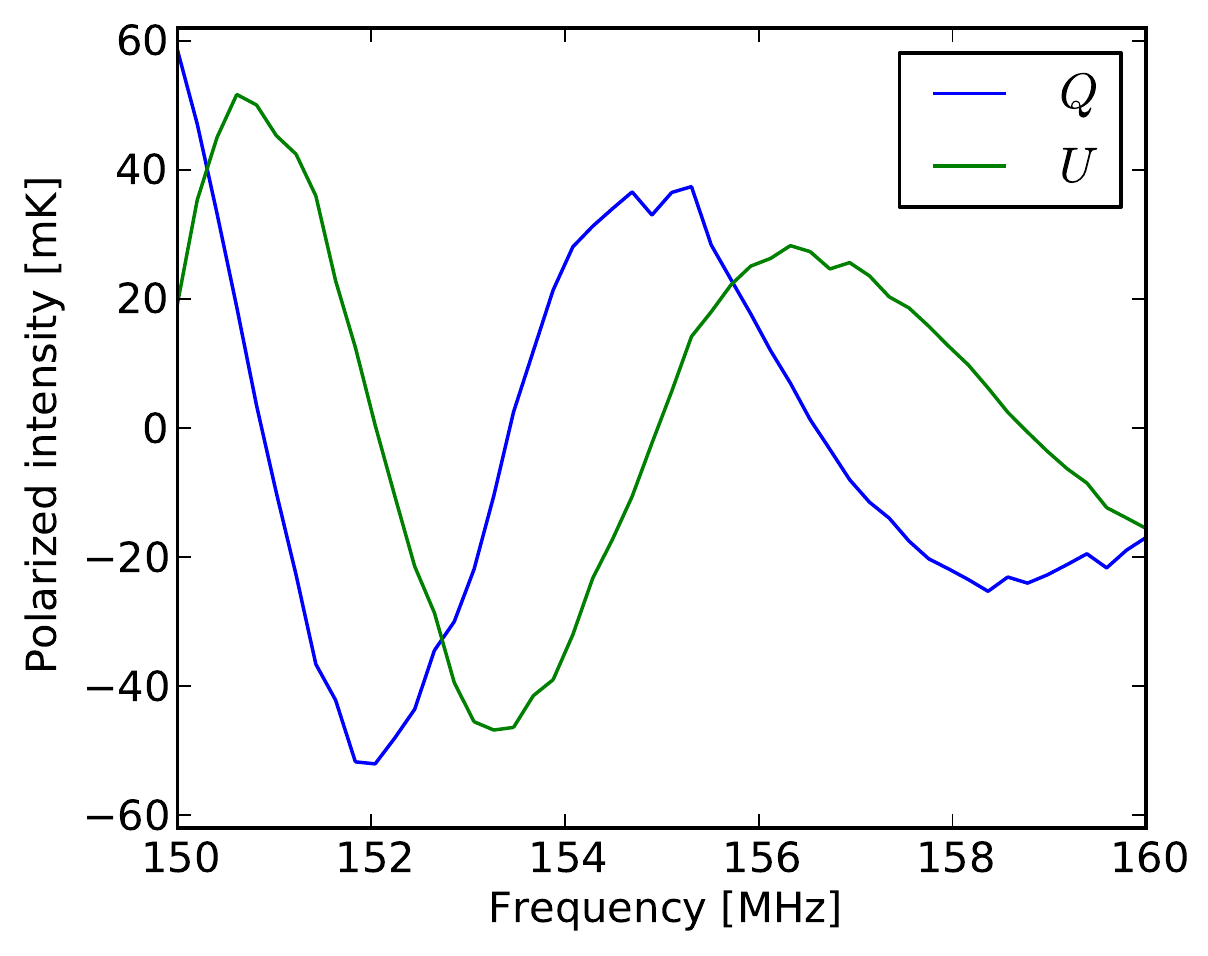}
\end{minipage}
\caption{\textit{Top}: Galactic diffuse foreground model in Stokes $Q$ and $U$ at 155 MHz created from the `model B' of \citet{je10}.
\textit{Bottom left}: The Power spectrum of the above model.
\textit{Bottom right}: Frequency spectra of the Stokes $Q$ and $U$ emission of the model for a single pixel.}
\label{f:model}
\end{figure*}

\begin{table}
\centering
\begin{minipage}{\linewidth}
\centering
\caption{Parameters related to the simulated observations:}
\label{t:3model}
\begin{tabular}{@{}lllr@{}}
\hline
\hline
\\
& NCP & 3C196 \\
\cline{2-3}\\
Phase center ($\alpha,\delta$) & 0$^\circ$, 90$^\circ$ & 123.4$^\circ$, 48.2$^\circ$ \\
Observing time & 13 hours & 8 hours \\
Baseline range & \multicolumn{2}{c}{30--180 $\lambda$} \\
Frequency range & \multicolumn{2}{c}{150--160 MHz} \\
Spectral subbands & \multicolumn{2}{c}{50} \\
Bandwidth & \multicolumn{2}{c}{10 MHz} \\
Spectral resolution & \multicolumn{2}{c}{0.195 MHz} \\
Integration time & \multicolumn{2}{c}{120 s} \\
Pixel scale & \multicolumn{2}{c}{18.75 arcmin} \\
\\
\hline
\hline
\end{tabular}
\end{minipage}
\end{table}

The results of this exercise are presented in terms of PS.
Some PS of polarized emission and leakage have been presented in A15, and in Section \ref{s:obs} of this paper.
However, they were calculated from observed images, whereas here we calculate PS directly from the simulated visibilities.
One of the motivations behind the current approach is that the sky model used here has very low resolution and we do not intend to gain any extra information from the images.
Also, it would be easier to characterize instrumental effects in visibility space, as we will not need to worry about imaging artifacts.
The procedure of creating PS from visibilities and the figures of merit used to present the final results are described in section \ref{s:vis2ps}.

\subsection{Model of diffuse polarized emission}\label{s:model}
We use the foreground simulations of \citet{je10}, in which the emission coefficients of the Galactic synchrotron and free-free emission are obtained given the cosmic-ray and thermal electron densities, and a Galactic magnetic field model.
It is assumed that both cosmic-ray and thermal electrons are mixed in a region of 1 kpc in depth along the LOS.
The synchrotron emission produced by the cosmic-ray electrons are depolarized due to differential Faraday rotation.
This is the `Model B' of \citet{je10} and we refer the readers to that paper for more details.

We have created 3D cubes of dimension $50^3$ for Stokes $Q$ and $U$ models within a field of $15^\circ \times 15^\circ$ and a frequency range of 150--160 MHz.
An example slice of the model is shown in Fig. \ref{f:model}.
Note that level of polarized foreground in this simulation is lower than the typical observed polarized emission, as shown in Fig. \ref{f:ps2d}.
However, the main result of this exercise will not be affected by this unrealistic choice of model, because we are only interested in the \textit{fractional} leakage.
The top panels of Fig. \ref{f:model} show Stokes $Q$ and $U$ emission at 150 MHz.
The power spectra of $P=Q+iU$ of this slice is shown in the bottom-left panel.
The high level of power at small $k$-scales is caused by the large-scale diffuse emission.
The frequency profile of an example pixel of this cube is shown in the bottom-right panel of the figure.
The figure shows spectral fluctations of the polarized emission caused by differential Faraday rotation of synchrotron emission by the mixed thermal and cosmic-ray electrons in the intervening medium.

\citet{je10} noted if a fraction of this polarized emission is leaked into total intensity, the leakage might mimic the EoR signal.
Since then, it has been found that in the 3C196 field, differential Faraday rotation is small and the observed diffuse polarized emission does not fluctuate too much along frequency \citep{je15}.
However, in Section \ref{s:obs}, we have shown that the situation is different in the NCP field.
Although the level of diffuse polarized emission in the NCP field is lower than that of the 3C196 field, the former has more fluctuation along frequency, i. e. it has higher power at high $k_\parallel$-scales or equivalently at high Faraday depths.
It would be useful to examine the effect of leakage on the EoR window in the worst case scenario, namely when the fluctuation of the emission along frequency is high.
Therefore, even though the choice of model is not important for the measurement of fractional leakage---as leakage is caused solely by $uv$-plane effects---, we have used a spectrally fluctuating model.

\subsection{Power spectrum from visibilities} \label{s:vis2ps}
We have used Stokes visibilities to create cylindrically averaged (2D) PS.
Consider a Stokes visibility $V_Z(b,t,\nu)$ for the baseline of length $b=\sqrt{u^2+v^2}$ at the time $t$ and frequency $\nu$, where $Z=I,Q,U,V$ denotes different Stokes parameters.
In synthesis observations, the position of a baseline with respect to the astronomical source changes as the Earth rotates, producing many more baselines than is possible with a snapshot observation.
In other words, if there are $n_b$ physical baselines, then after synthesizing over $n_t$ timesteps the total number of baselines will be $n_b\times n_t$.
As the baselines created by the synthesis can be considered independent baselines in themselves, we can re-write $V_Z(b,t,\nu)$ as just $V_Z(b,\nu)$ where $b$ is the length of any of the $n_b\times n_t$ baselines.

To produce PS, the $uv$-plane within the baseline range $b_{min}$--$b_{max}$ is gridded\footnote{Visibility gridding is done using the {\tt EXCON} (\url{https://sourceforge.net/projects/exconimager}) imager created by Sarod Yatawatta.} in $N_b\times N_b$ pixels.
The width and height of each pixel is $16\lambda$, corresponding to the diameter of a LOFAR core HBA-station.
To correct for the $w$-terms, the visibilities were also $w$-projected using 32 $w$-planes.
A visibility cube is produced taking the gridded visibilities for all spectral subbands.
The visibility in each voxel of this cube can be written as $V_Z(b_n,\nu)$, where $b_n$ refers to the $n$-th pixel, $n$ going from $0$ to $(N_b-1)^2$.

Each pixel of the visibility grid is Fourier transformed along the frequency axis using the one-dimensional FFT algorithm.
The Fourier conjugate of frequency is `delay' ($\tau$) and the transform can be written mathematically as
\begin{equation}
\tilde{V}_Z(b_n,\tau) = \frac{1}{N_\nu} \sum_{\nu=0}^{N_\nu-1} V_Z(b_n,\nu) \exp{\left[-2\pi i \frac{\nu\tau}{N_\nu}\right]}
\end{equation}
where $N_\nu$ is the number of frequency channels.
The squared absolute values of $\tilde{V}_Z(b_n,\tau)$ give the 3D PS, i. e. $P_{3D}(b_n,\tau)=|\tilde{V}_Z(b_n,\tau)|^2$.

To create 2D PS, the $uv$-plane of $P_{3D}$ is divided into $N$ annuli and the powers within an annulus are averaged for each delay.
The averaged power at a certain delay and a certain $uv$-annulus $b_N$
\begin{equation}
P_{2D}(b_N,\tau) = \frac{1}{N_p} \sum_{b=b_N}^{b=b_{N+1}} P_{3D}(b,\tau)
\end{equation}
where $N_p$ is the number of pixels within the annulus $b_N$.

Delay $\tau$ is related to $k_\parallel$ \citep[equation 6]{Thyagarajan2015a}, and the baseline length corresponding to a particular annulus $b_N$ is related to $k_\perp$ (A15, equation 30).
We have presented $P_{2D}(b_N,\tau)$ in terms of $k_\perp$ and $k_\parallel$, i. e. as $P_Z \equiv P_{2D}(k_\perp,k_\parallel)$.
From here, dimensionless power spectrum is calculated as $\Delta^2(k_\perp,k_\parallel) = k_\perp^2 k_\parallel P_Z(k_\perp,k_\parallel) / (2\pi)^2$.

We are interested in Stokes $Q,U$ and their leakage into Stokes $I$.
The power of Stokes $I,Q,U$ visibilities are denoted by $P_I,P_Q,P_U$, respectively, and the power of the linear polarization $Q+iU$ by $P_P$.
Then, the fractional leakage from $Q,U$ to $I$
\begin{equation} \label{eq:LI}
L_I = \sqrt{\frac{P_I}{P_P}}\times 100.
\end{equation}
$L_I$ represents the leakage into Stokes $I$ as a percentage of linear polarization as a function of $k_\perp$ and $k_\parallel$.
Both the 2D spectrum and histogram of $L_I$ for different diameters and toward different directions are presented below.

\subsection{Results}
Cylindrically averaged 2D PS of the simulated polarized emission observed by LOFAR and its leakage into total intensity are presented in Fig. \ref{f:3C196_ps} and \ref{f:NCP_ps}.
Fig. \ref{f:3C196_ps} shows the spectra for the 3C196 field, and Fig. \ref{f:NCP_ps} for the NCP field.
In both figures, the columns represent $P_P$, $P_I$ and $L_I$ spectra from left to right, respectively.
And the rows represent the spectra within the diameters of $15^\circ$, $9^\circ$ and $4^\circ$ from top to bottom, respectively.

\subsubsection{Polarization power}
\begin{figure*}
\centering
\includegraphics[width=0.75\linewidth]{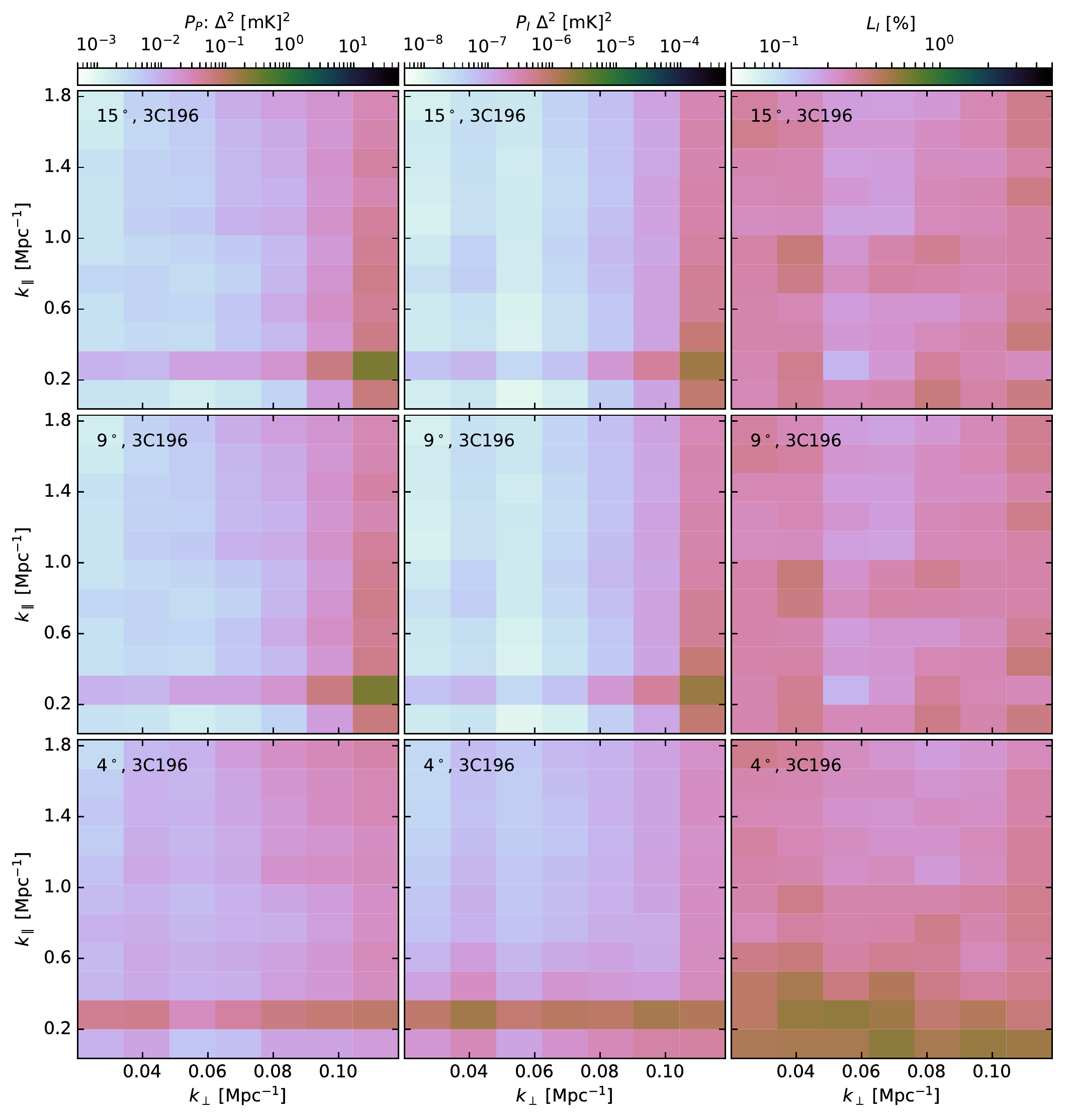}
\caption{Power spectra created from the visibilities corresponding to the model of Fig. \ref{f:model} toward the 3C196 field. The three rows represent three fields of view centered around 3C196: from top to bottom $15^\circ \times 15^\circ$, $9^\circ \times 9^\circ$ and $4^\circ \times 4^\circ$ deg$^2$ respectively.
The three columns represent Stokes $|Q+iU|\equiv P$, $I$, and $\sqrt{I/P}$ from left to right respectively.}
\label{f:3C196_ps}
\end{figure*}

\begin{figure*}
\centering
\includegraphics[width=0.75\linewidth]{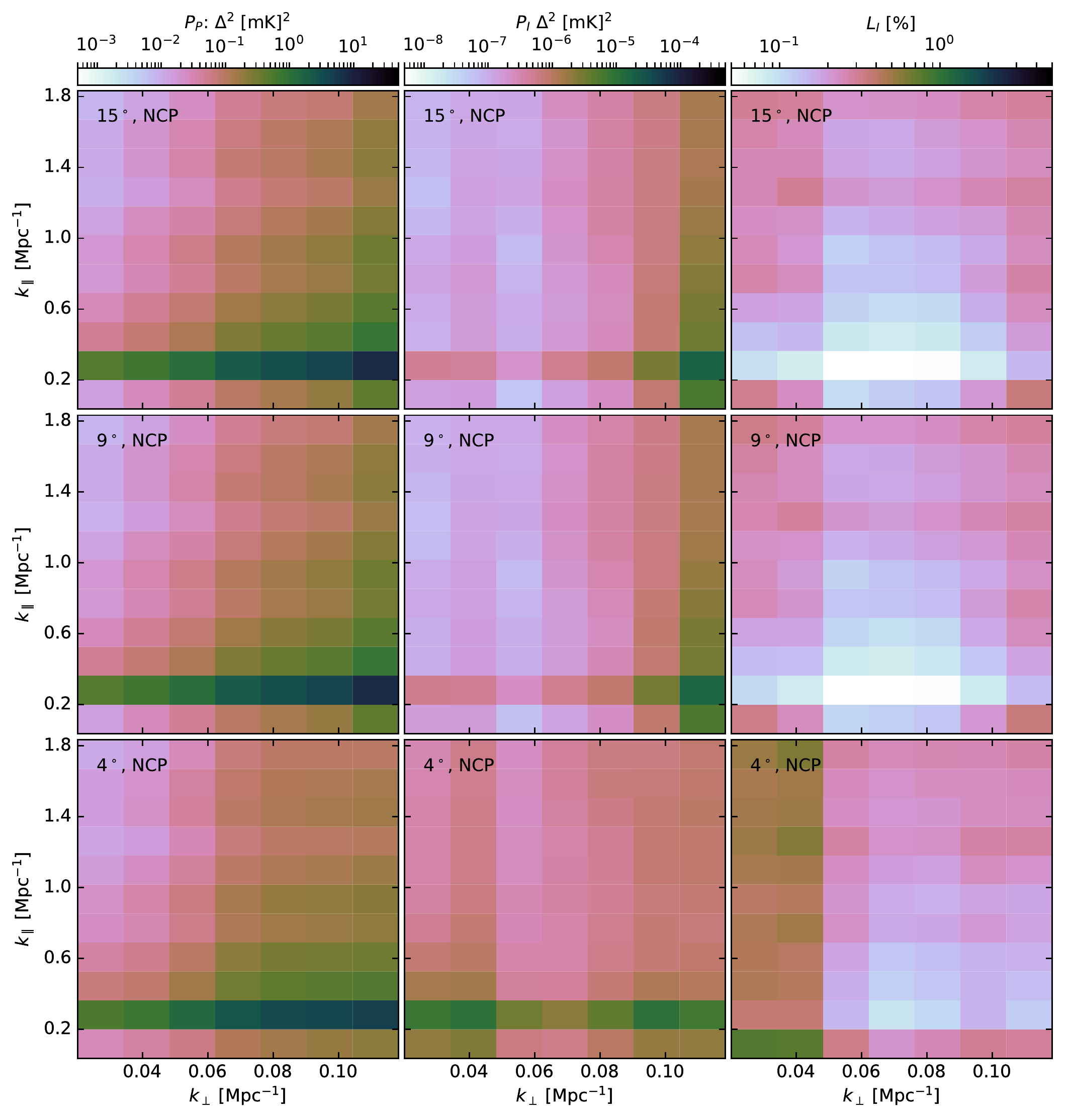}
\caption{Power spectra created from the visibilities corresponding to the model of Fig. \ref{f:model} toward the NCP field. The three rows represent three fields of view centered around NCP: from top to bottom $15^\circ \times 15^\circ$, $9^\circ \times 9^\circ$ and $4^\circ \times 4^\circ$ deg$^2$ respectively.
The three columns represent Stokes $|Q+iU|\equiv P$, $I$, and $\sqrt{I/P}$ from left to right respectively.}
\label{f:NCP_ps}
\end{figure*}

PS of linear polarization ($P_P$) show the characteristics of the sky model in both fields.
In the left panels of Fig. \ref{f:3C196_ps} and \ref{f:NCP_ps}, most polarized power are found within a wedge shaped region at high-$k_\perp$, but the the wedge is located at a higher $k_\parallel$ scale as compared to the observed emission of Fig. \ref{f:ps2d}.
This high-$k_\parallel$ power is, of course, due to the spectral nature of the model diffuse emission shown, e. g., in the bottom-right panel Fig. \ref{f:model}.
A comparison between Fig. \ref{f:3C196_ps} and \ref{f:NCP_ps} shows that more polarization power is predicted in the NCP than in the 3C196 field.
Although the simulated sky model is same for both fields, the output visibilities are different because the visibilities have been convolved with two different PBs, and also the fields have been observed for different durations of time---3C196 for 8 hours and NCP for 13 hours.

\begin{figure*}
\centering
\includegraphics[width=0.48\linewidth]{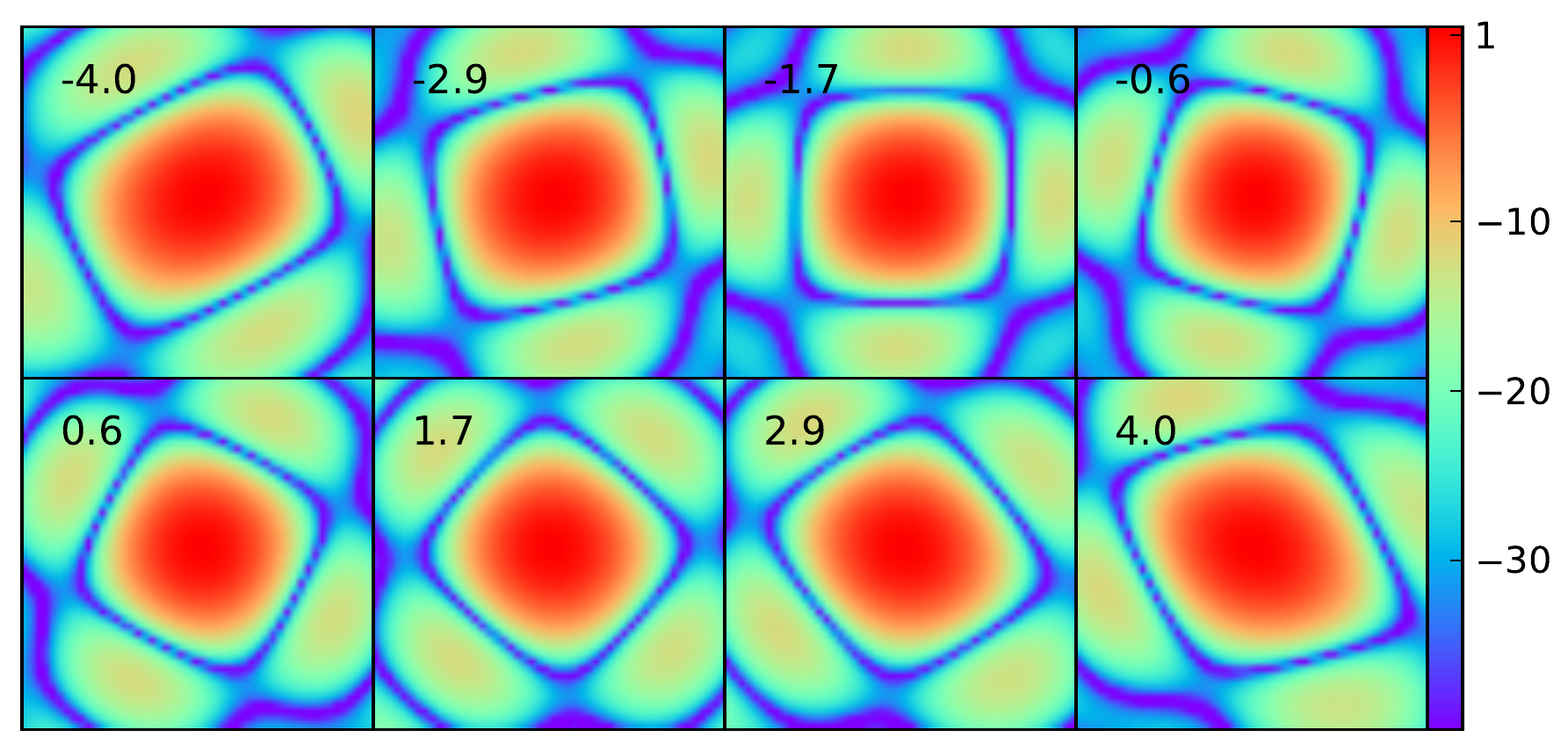}
\includegraphics[width=0.48\linewidth]{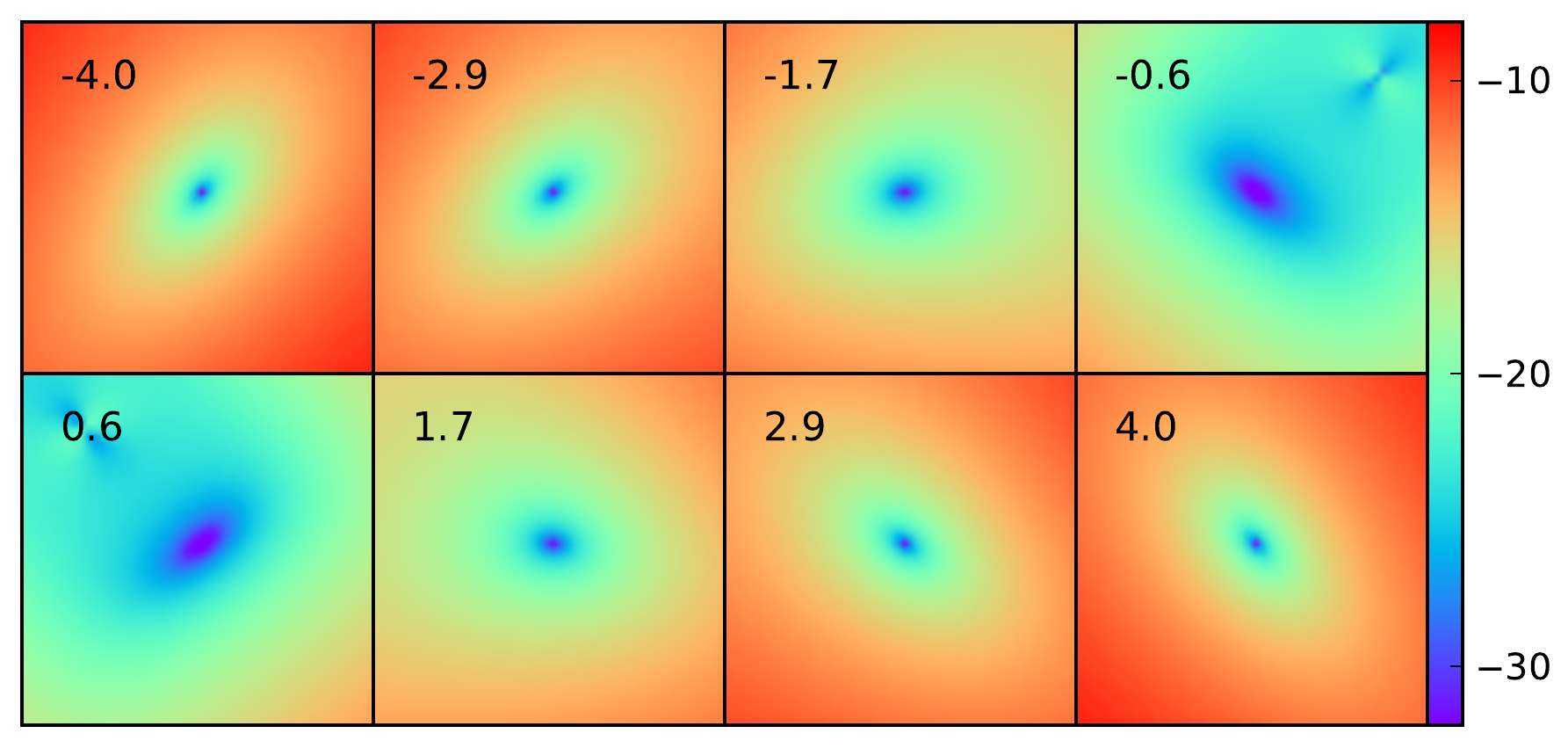}

\includegraphics[width=0.73\linewidth]{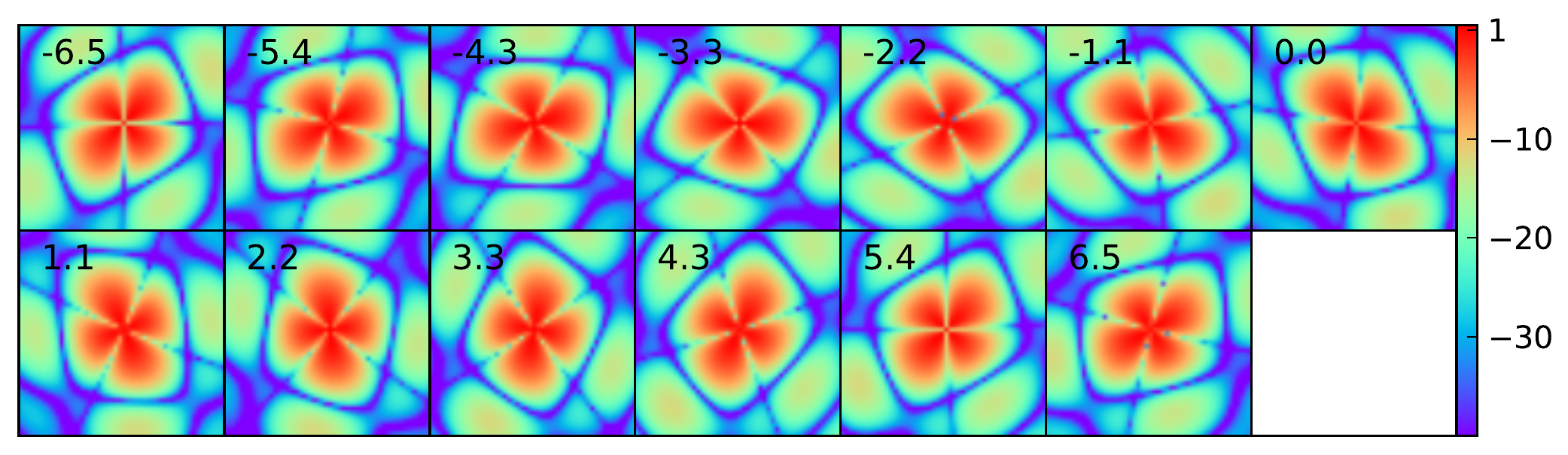}
\includegraphics[width=0.73\linewidth]{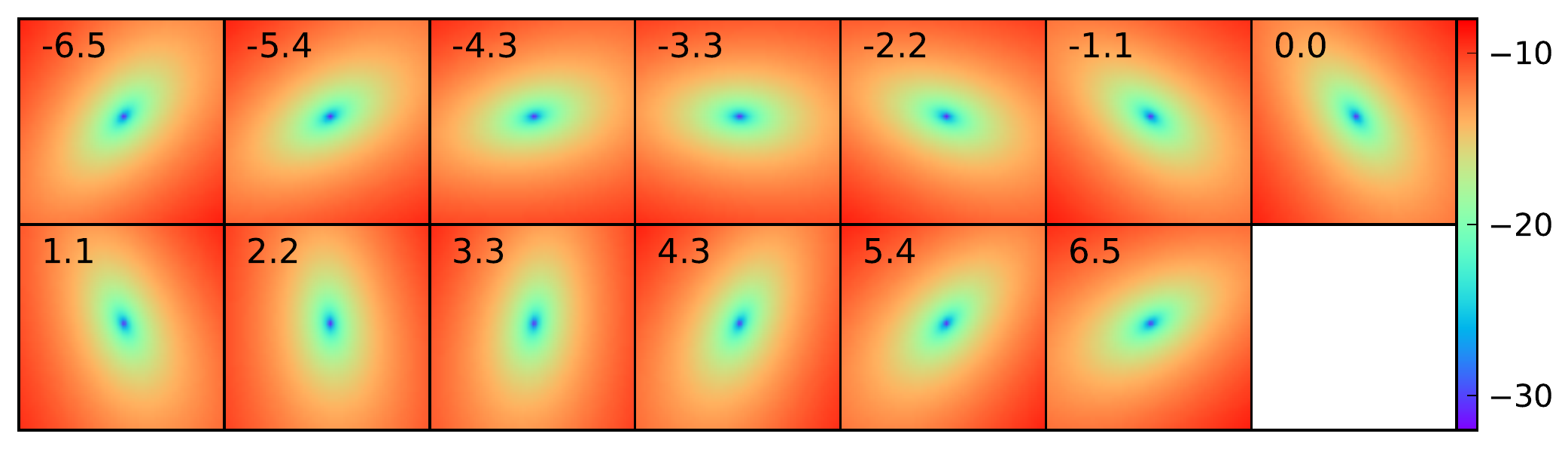}

\caption{\textit{Top left}: LOFAR PB model for linear polarization within $15^\circ\times 15^\circ$ of the \textit{3C196 field} for a single baseline at 150 MHz, i. e. $\sqrt{M_{22}^2+M_{33}^2}$ where the subscripts represent the components of the PB Mueller matrix.
\textit{Top right}: Leakage from linear polarization to Stokes $I$ as a fraction of Stokes $I$, i. e. $\sqrt{M_{12}^2+M_{13}^2}/M_{11}$, for the same field.
The \textit{middle} and \textit{bottom} rows show the same polarization PB model and leakage terms for the \textit{NCP field}.
All PBs have same FoV, and are normalized with respect to the Mueller matrix at the phase center.
The colorbars are shown in dB unit, and the numbers on the top-left corners show the hour angles at which the PBs are measured.}
\label{f:beams}
\end{figure*}

The difference between the polarization PBs toward the 3C196 and NCP fields can be seen in the top-left and middle rows of Fig. \ref{f:beams}.
The figures show some components of the $4\times 4$ PB Mueller matrix (e. g. A15, fig. 2a), the outer product of the PB Jones matrices of two stations constituting a baseline.
$M_{22}$ (signifying second row, second column) and $M_{33}$ components of the matrix represent the Stokes $Q$ and $U$ PBs, respectively.
The aforementioned figures represent the polarization PB $\sqrt{M_{22}^2+M_{33}^2}$ within a diameter of $15^\circ$ normalized with respect to the phase center for the 3C196 (top-left) and NCP (middle) fields.
We see that the shape of the polarization PB is different in the two fields---the latter has a 2-fold symmetry, whereas the former is either circular or elliptical depending on the hour angle.
Different panels in the figures show PBs at different hour angles, and we can see the rotation of the PB with the apparent rotation of the sky.
The 3C196 field rises, reaches very close to the zenith and then sets during an observation, but the NCP field continually rotates around the north celestial pole with the rotation of the Earth.

The PS of Fig. \ref{f:3C196_ps} and \ref{f:NCP_ps} show the extent of contamination of the `EoR window' by the leakage of a spectrally unsmooth polarized emission.
Looking at the middle columns of Fig. \ref{f:3C196_ps} and \ref{f:NCP_ps}, one can clearly see that there is significant foreground power at higher $k_\parallel$ scales contaminating the expected EoR signal.
Note that most diffues emission are still concentrated near $k_\parallel\sim 0.2$, because the frequency of the spectral variation seen in Fig. \ref{f:model} (bottom-right panel) is still quite low.
Comparing the simulated leakages in the two fields, we can see that the contamination of the EoR window is higher in `NCP' than in `3C196'.
However, in real observations, this level of spectral structure has not been found in all fields.
For example, as mentioned in Section \ref{s:obs}, between the observed emission in the 3C196 and NCP fields, the former shows limited spectral fluctuation and almost no leakage in the EoR window, whereas the latter suffers from considerable level of leakage in the EoR window due to its high-rotation measure polarized emission.

\subsubsection{Fractional leakage power and rms}
The PS of the leakage of polarized emission into Stokes $I$ ($P_I$) are shown in the middle panels of Fig. \ref{f:3C196_ps} (3C196 field) and \ref{f:NCP_ps} (NCP field).
The figures show that $P_I$ is almost a scaled down version of $P_P$ within any diameter and toward both the observing fields.
The square-root of the ratio between the two PS ($L_I$) are shown in the right-hand panels of both figures as a percentage.
Because $L_I$ is the square root of power or variance, it represents the \textit{rms} of the fractional leakage as a function of $k_\perp$ and $k_\parallel$.

In both 3C196 and NCP fields, $L_I$ varies little over the instrumental $k$-space.
Whatever variation is present is mostly due to the sample variance.
The difference among the three areas $15^\circ\times 15^\circ$, $9^\circ\times 9^\circ$ and $4^\circ\times 4^\circ$ is not significant, although the fraction seems to be higher within $4^\circ\times 4^\circ$ compared to the other two.
There is significant difference between the $L_I$ PS of the 3C196 and NCP fields; it is considerably lower in the NCP field.

\begin{figure*}
\centering
\includegraphics[width=0.8\linewidth]{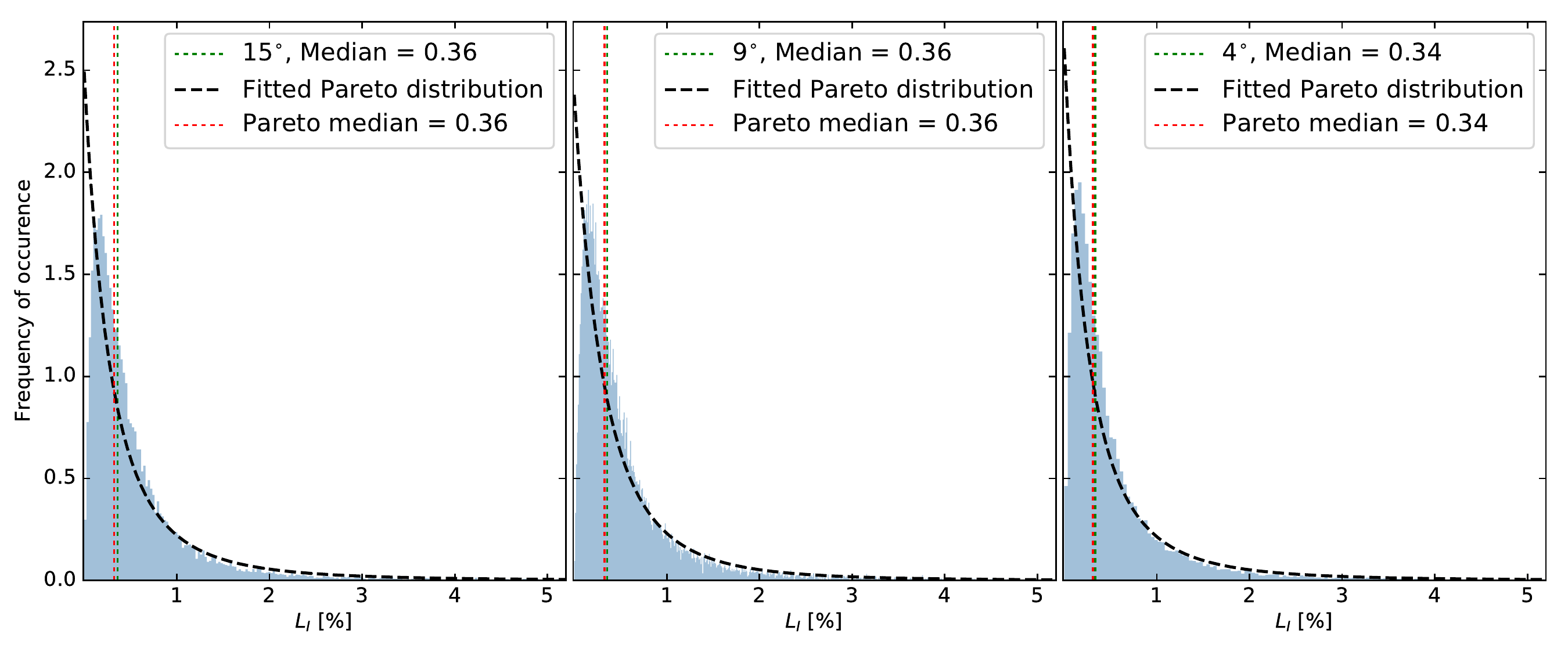}
\includegraphics[width=0.8\linewidth]{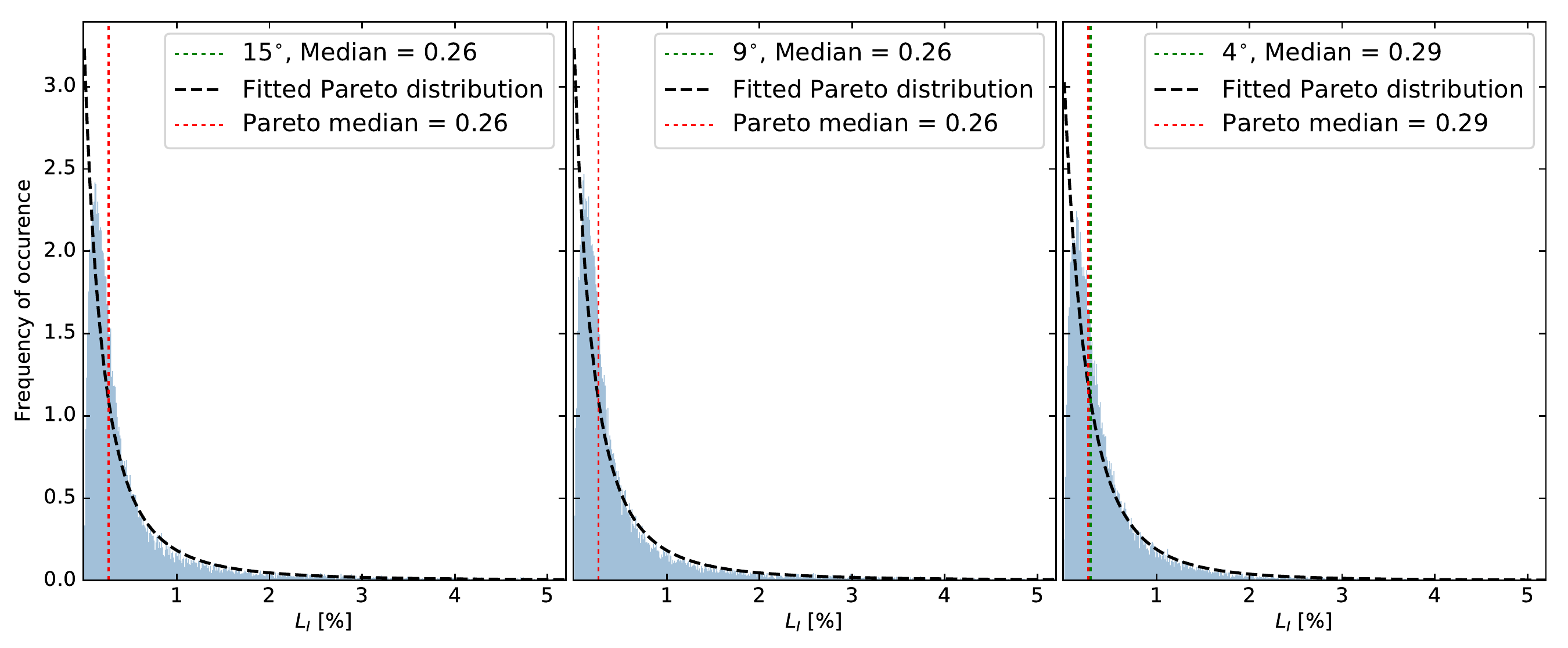}
\caption{Histograms of the leakage RMS ($L_I^{\tt rms}$) over the instrumental $k$-space of the 3C196 (top row) and NCP (bottom row) fields.
The three columns correspond to the fields of view of $15^\circ \times 15^\circ$, $9^\circ \times 9^\circ$ and $4^\circ \times 4^\circ$ from left to right respectively.
The median values calculated from the original data and the fitted Pareto distribution are shown on the insets.}
\label{f:hist}
\end{figure*}

The comparison between the two fields and the three areas would be easier if we look at the histogram of $L_I$ (Fig. \ref{f:hist}) for all voxels in the cube before averaging cylindrically or spherically.
The top and bottom rows of the figure show the histograms for the 3C196 and NCP fields, respectively.
The three columns show the histograms for the three areas.
The mean, median, standard deviation and position of the peak of $L_I$ for all these distributions are presented in Table \ref{t:hist-3C196} and \ref{t:hist-NCP}.
The histograms have several general characteristics: they all have a long tail, their medians are close to the positions of the peaks, and the means are greater than the medians due to the tails.
Therefore, median would be a better estimator for $L_I$.

\begin{table}
\centering
\caption{Statistics of $L_I$ (\%) for different fields of view in the 3C196 field.}
\label{t:hist-3C196}
\begin{tabular}{l|lll|}
\hline
\multicolumn{1}{|l|}{Fields of view}     & \multicolumn{1}{c}{$15^\circ$} & \multicolumn{1}{c}{$9^\circ$} & \multicolumn{1}{c|}{$4^\circ$} \\ \hline

\multicolumn{1}{|l|}{Mean} & 0.62 & 0.61 & 0.59 \\
\multicolumn{1}{|l|}{Median} & 0.36 & 0.36 & 0.34 \\
\multicolumn{1}{|l|}{Standard deviation} & 1.39 & 1.12 & 1.47 \\
\multicolumn{1}{|l|}{Peak} & 0.18 & 0.15 & 0.15 \\ \hline

\end{tabular}
\end{table}

\begin{table}
\centering
\caption{Statistics of $L_I$ (\%) for different fields of view in the NCP field.}
\label{t:hist-NCP}
\begin{tabular}{l|lll|}
\hline
\multicolumn{1}{|l|}{Fields of view}     & \multicolumn{1}{c}{$15^\circ$} & \multicolumn{1}{c}{$9^\circ$} & \multicolumn{1}{c|}{$4^\circ$} \\ \hline

\multicolumn{1}{|l|}{Mean} & 0.63 & 0.63 & 0.49 \\
\multicolumn{1}{|l|}{Median} & 0.26 & 0.26 & 0.29 \\
\multicolumn{1}{|l|}{Standard deviation} & 1.76 & 1.67 & 0.94 \\
\multicolumn{1}{|l|}{Peak} & 0.09 & 0.09 & 0.12 \\
\hline
\end{tabular}
\end{table}

The median of the rms fractional leakage is higher in the 3C196 field compared to the NCP, which seems paradoxical given that the PB model shows more leakage in the NCP field, as evident from the top-right and bottom rows of Fig. \ref{f:beams}.
Leakage is lower near the zenith and increases as one moves away from the zenith.
Because the 3C196 field comes very close to the zenith during the mid-point of the observation, the leakage decreases substantially.
But the NCP field rotates around a fixed elevation resulting in a constant leakage pattern throughout the observing time.
Therefore, from the PB plots the 3C196 field seems to exhibit more leakage.
However, in the PS, the leakages at all hour angles are averaged, and we think Stokes $Q$ and $U$ are averaged down more in the NCP than in the 3C196 field resulting in a lower rms fractional leakage in the former field.

The median of $L_I$ has a very weak dependence on the area of the field; it remains almost constant.
This also seems paradoxical at first glance, because the fractional leakage calculated directly from the model PB increases as a function of distance from the phase center (A15, fig. 2b).
However, as we go away from the phase center, the fractional leakage increases, but the power of the polarized emission decreases.
This can be clearly seen by comparing the top-left and middle rows (representing the co-polarization or diagonal elements of the Mueller matrix) with the top-right and bottom rows (cross-polarization or off-diagonal terms) of Fig. \ref{f:beams}.
The co-pol terms attenuate the polarized emission, and the cross-pol terms determine the level of leakage.
In this case, these two effects are canceling each other out so that the rms fractional leakage does not vary as one goes farther away from the phase center.

\subsubsection{Statistics of fractional leakage}
Radio astronomical signals are random noises and follow Rayleigh, or Gaussian, distributions.
In our case, $Q$ and $U$ are Gaussian distributed complex numbers, so $|Q+iU|$ follows a Rayleigh distribution.
Furthermore, Stokes $I$ is a Gaussian distributed complex number, making its absolute value Rayleigh distributed.
Therefore, fractional leakage $L_I$, defined in Eqn. \ref{eq:LI}, would follow the distribution of the ratio of two Rayleigh distributions.
The probability density function (pdf) of a Rayleigh distribution
\begin{equation}
f(x;\sigma) = \frac{x}{\sigma^2} \ e^{-x^2/2\sigma^2}
\end{equation}
where $\sigma$ is the scale parameter, and the pdf of the ratio of two such distributions \citep{Shakil2011}
\begin{equation}
f(r;\sigma_1,\sigma_2) = \frac{2\sigma_1^2\sigma_2^2 r}{(\sigma_2^2 r^2+\sigma_1^2)^2}
\end{equation}
where $\sigma_1$ and $\sigma_2$ are the scale parameters of the two distributions.
$f(r;\sigma_1,\sigma_2)$ is equivalent to the pdf of the square root of the Pareto distribution with shape parameter 1.
Therefore, fitting the $L_I$ distributions shown in Fig. \ref{f:hist} with a Pareto distribution would give a more stringent estimate of the fractional leakages toward the 3C196 and NCP fields.

The black dashed lines in all the panels of Fig. \ref{f:hist} show the pdf of the fitted\footnote{The fitting was performed using the python statistical function {\tt scipy.stats.pareto} with a shape parameter 1.} Pareto distributions.
Curiously, the medians of the fitted distributions match with the medians calculated directly from the data, as shown in the inset of each panel.
Therefore, one can take the medians of the data or the fitted distributions as the final estimate of the rms fractional leakage, $L_I$.
Because this value does not vary significantly with the diameter of the field, we take the average of the medians within the three diameters, for each field.
The resulting rms fractional leakage for the 3C196 and NCP fields are $0.27\%$ and $0.35\%$, respectively.

\section{Discussion}
The observed polarized emission is lower in amplitude but has larger fluctuation along frequency in the NCP field, compared to the 3C196 field because in the former field emission is present over a much wider range of Faraday depths.
The power of the polarized emission ($P_P$) in both fields has been shown in Fig. \ref{f:ps2d} as cylindrically averaged PS.
A fraction of this power leaks into Stokes $I$ because of the polarized PB.
The fractional leakage ($L_I$) has been calculated through simulations within three different diameters for the two fields.
The observed $P_P$ and model $L_I$ (according to the model PB) can be used to calculate the rms of the leakage into Stokes $I$ that should be expected in the two fields.
Mathematically, the rms of the expected leakage
\begin{equation} \label{eq:LIexp}
L_I^{\tt exp} = \sqrt{P_P^{\tt obs}} \times L_I^{\tt model}.
\end{equation}
We have seen that $L_I$ varies very little over the instrumental $k$-space, and hence the leakage PS can be considered to be an almost scaled-down version of the polarization PS.
The same scaling relationship is also seen seen in Fig. 12(e) of A15.
$L_I^{\tt model}$ is $0.27\%$ for the NCP field, $0.35\%$ for the 3C196 field, as estimated by fitting Pareto distributions to the histograms of $L_I$ above.

\begin{figure}
\centering
\includegraphics[width=0.9\linewidth]{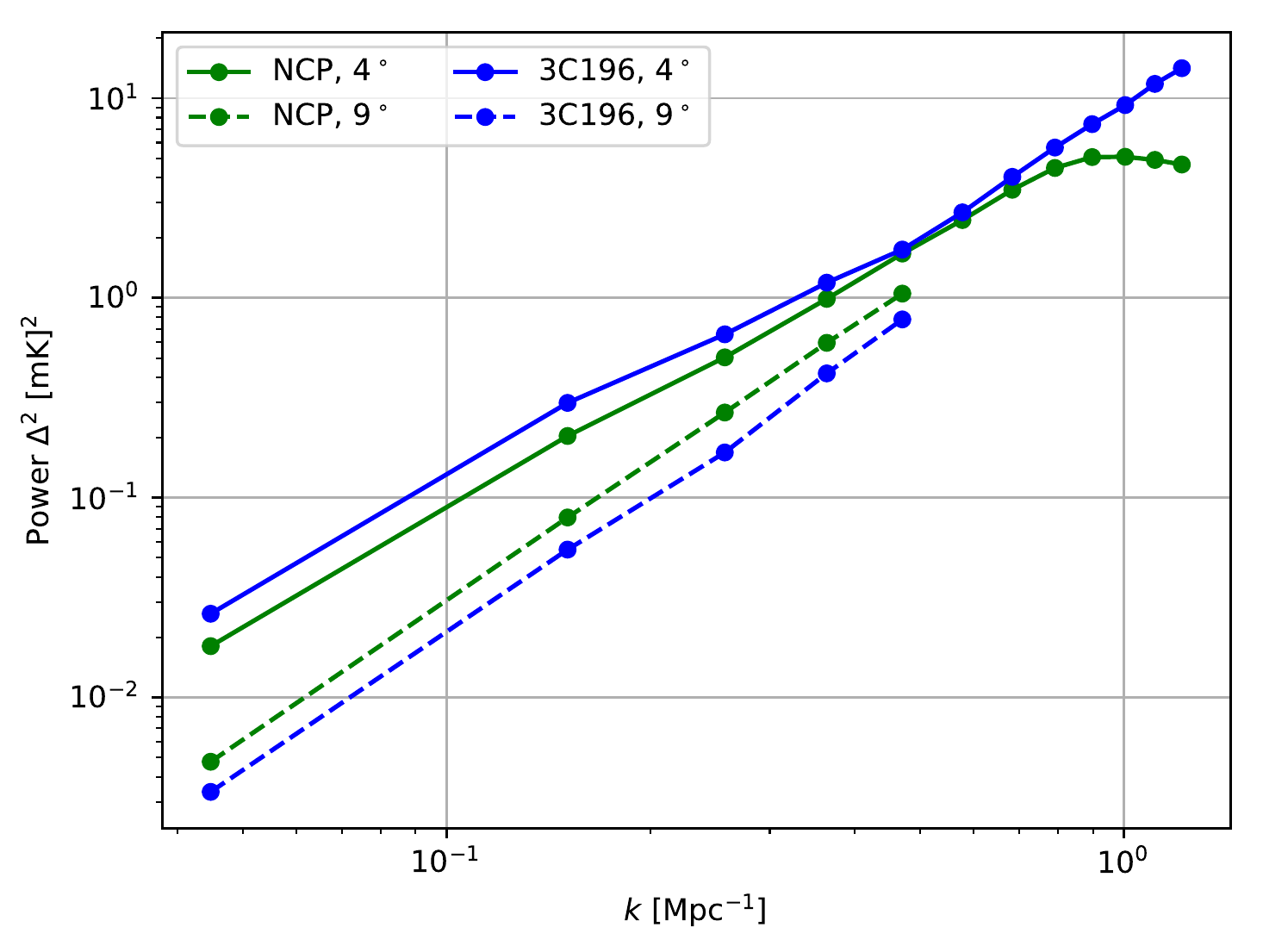}
\caption{Spherically averaged power spectrum of the expected polarization leakage into Stokes $I$ ($[L_I^{\tt exp}]^2$) in the 3C196 (blue) and NCP (green) fields within $4^\circ\times 4^\circ$ (solid) and $9^\circ\times 9^\circ$ (dashed).}
\label{f:expL}
\end{figure}

In Fig. \ref{f:expL}, we show the variance of the expected leakage, $[L_I^{\tt exp}]^2$, in the 3C196 (blue lines) and NCP (green) fields within a diameter of $4^\circ$ (solid) and $9^\circ$ (dashed).
The figure shows that leakage is lower in the $9^\circ\times 9^\circ$ fields because they are more noise-dominated compared to the smaller fields.
The leakage in the NCP field is lower than that of the 3C196 field within a diameter of $4^\circ$, but the situation is reverse within a diameter of $9^\circ$.
The reversal has nothing to do with the fractional leakage, as we have used same $L_I$ for both diameters, and thus probably shows that the noisy polarized emission within the larger diameter is averaged down more in the 3C196 field.
As noted earlier, the NCP field exhibits higher power at high $k_\parallel$ in the 2D PS, and hence even if the expected level of leakage is similar in both fields, the `EoR window' would suffer more from leakage-contamination in the NCP field.

One limitation of our results regarding $L_I$ is that we do not know the accuracy of the LOFAR PB model very well outside the first null, which has a diameter of $6.4^\circ$ at 150 MHz.
\citet{as16} found that the PB model has $10\%$ error in predicting polarization leakage within the first null.
We do not expect the same level of accuracy outside the field of view, but more work is needed to quantify this (future holographic measurements of the LOFAR beam will be helpful in this regard).
In this paper, the PB model out to a maximum diameter of $15^\circ$ has been used, and we are bound to be affected by PB model errors to some extent.

However, the general formalism of characterizing the fractional leakage described here is not prone to the specific model errors of our simulations.
If one knows the PS of the low-frequency extragalactic and Galactic polarized emission and the full-polarization wide-bandwidth beamshapes of the antennae of an array, one would be able to predict the PS of the polarization leakage into Stokes $I$ using Eqn. \ref{eq:LIexp}.
In this paper, we have not found any dependence of $L_I$ on the diameter of the observing field in case of LOFAR.
However, this is only because LOFAR antennae have narrower beamwidth compared to, e. g., those of PAPER, HERA or MWA.
In the case of an antenna with wider beamwidth, the decrease of power might not be enough to suppress the increase of leakage as a function of distance from the phase center.

The mechanism of predicting polarization leakage shown here alludes to a potential way of correcting the total intensity PS for the leakage bias, in a similar fashion as the correction for the noise bias.
For example, in case of LOFAR, PS of the extragalactic and Galactic polarized emission can be created from observations, then the leakage PS can be predicted using our simulation tools, and finally this leakage PS can be subtracted from the total intensity PS, thereby cleaning off leakage from the EoR window to a first order.

\section{Conclusion}
In this paper, we have presented cylindrically and spherically averaged power spectra (PS) of the observed Galactic diffuse polarized emission in the 3C196 and NCP fields of the LOFAR-EoR key science project.
The PS have been produced from an 8-hour synthesis observation of the 3C196 field, and a 13-hour observation of the NCP field for 50 spectral subbands ranging from 150 to 160 MHz.
A version of the 3C196 PS was presented in \citet{as15}, but unlike that paper, here we have created the PS directly from the observed image cubes without re-convolving them with the PB.
The main difference between the PS of the two fields seen in Fig. \ref{f:ps2d} is that, in the 3C196 field polarization power is restricted within a `wedge' at low-$k_\parallel$ and high-$k_\perp$, whereas in the other field power is more distributed and one can see a considerable amount of power at the high-$k_\parallel$, low-$k_\perp$ corner of the PS, i. e. in the `EoR window'.
This is because the diffuse emission in latter field has more spectral fluctuations.
The PS were produced within two different diameters: $4^\circ$ and $9^\circ$.
Power slightly decreases within the larger area, because the region outside a diameter of $4^\circ$ is dominated by noise.

We have also determined the fraction of polarized power that would leak into Stokes $I$ PS of the two fields, contaminating the EoR window, within three different diameters: $15^\circ$, $9^\circ$ and $4^\circ$.
A leakage PS of the 3C196 field was presented in \citet{as15} by simulating a sky modeled from real observations of Galactic diffuse polarized emission.
To avoid the effect of noise, here we have measured the PS for both the 3C196 and NCP fields by simulating LOFAR observations of a model Galactic-diffuse-polarized emission created by \citet{je10}.
The model emission has significant fluctuations along frequency due to differential Faraday rotation, and the spectrally fluctuating emission contaminate the EoR window of the PS considerably.
The square root of the ratio of the leakage power and linear polarization power, i. e. rms fractional leakage ($L_I$), has been found to vary very little over the instrumental $k$-space.
Histograms of $L_I$ (Fig. \ref{f:hist}) show this more clearly.

The distributions of $L_I$ are fitted with Pareto distributions, and the medians of the fitted distributions, which agree very well with the medians calculated from the data directly, are taken to be the best estimates of $L_I$.
The most interesting result of this paper is that, the rms fractional leakage does not change because of the leakages from outside the first null of the PB.
This is due to the fact that, although leakage increases with distance from the phase center, polarized power decreases due to attenuation by the polarization PB, and the two cancel each other out.
The resulting rms fractional leakage has a median of $0.35\%$ in the 3C196 field, and $0.27\%$ in the NCP field.

After calculating the PS from observation, and the fractional leakages from PB model simulations, we have predicted the level of leakage to be expected in the EoR windows of the two fields within different diameters (see Fig. \ref{f:expL}).
The leakage into Stokes $I$ is lower within a larger diameter, because the region outside the beamwidth is dominated by noise.
Leakage in the 3C196 field is higher than the NCP field within $4^\circ$, but the situation is reversed within a diameter of $9^\circ$, showing that noisy polarized emission is averaged down more in the former field.

One limitation of this work is that we have used fields much wider than the width of the PB for our simulations, but we do not know the accuracy of the PB very well outside the first null.
But the basic formalism behind our analyses is not prone to this uncertainty.
We have basically shown that it is possible to predict the leakage power spectrum, which in turn could be subtracted from the total intensity power spectrum, thereby cleaning off leakage from the `EoR window'.
This \textit{removal} through bias correction might indeed be needed, if the leakage of polarized emission is found inside the EoR window, as we have seen in the NCP field here.
However, if the polarized emission is spectrally smooth, which is the case in the 3C196 field, \textit{avoiding} its leakage would be enough.

\section*{Acknowledgments}
KMBA and LVEK acknowledge the financial support from the European Research Council under ERC-Starting Grant FIRSTLIGHT -- 258942, which is the primary funding source for this research.
KMBA is currently supported by a Postdoctoral Fellowship from the Rhodes University Centre for Radio Astronomy Techniques \& Technologies and the Centre for Radio Cosmology at the University of Western Cape, jointly.
VJ acknowledges the financial support from The Netherlands Organization for Scientific Research (NWO) under VENI grant -- 639.041.336.
AGdB and VNP acknowledge support from the ERC (grant 339743, LOFARCORE).
Finally, thanks to Abhik Ghosh for useful suggestions.

\bibliographystyle{mnras}
\bibliography{ref}



\label{lastpage}

\end{document}